\documentclass[fleqn,usenatbib,useAMS]{rasti}
\usepackage{graphicx}  
\usepackage{amsmath}   
\usepackage{amssymb}   
\usepackage{mathptmx}  
\usepackage{lastpage}  
\usepackage{orcidlink} 
\usepackage{balance}   
\usepackage{etoolbox}  
\let\oldorcidlink\orcidlink
\renewcommand\orcidlink[1]{\smash{\raisebox{.3em}{\Large{\oldorcidlink{#1}}}}}
\newcommand{\secref}[1]{\hyperref[#1]{Section~\ref*{#1}}} 
\newcommand{\secreftwo}[2]{\hyperref[#1]{Sections~\ref*{#1}} and \ref{#2}} 
\newcommand{\figref}[1]{\hyperref[#1]{Fig.~\ref*{#1}}} 
\newcommand{\appref}[1]{\hyperref[#1]{Appendix~\ref*{#1}}} 
\newcommand{\tabref}[1]{\hyperref[#1]{Table~\ref*{#1}}} 
\newcommand{\eqreft}[1]{\hyperref[#1]{equation~(\ref*{#1})}} 
\newcommand{\eqrefp}[1]{\hyperref[#1]{equation~\ref*{#1}}} 

\title[The WEAVE acquisition and guiding software]{The WEAVE acquisition and guiding software: \\pattern recognition-based acquisition and multi-fibre guiding}

\author[Gafton et al.]{%
Emanuel Gafton\orcidlink{0000-0003-0781-6638},$^{1}$\thanks{E-mail:             
    \href{mailto:ega@ing.iac.es}{ega@ing.iac.es} (EG); \href{gavin.dalton@physics.ox.ac.uk}{gavin.dalton@physics.ox.ac.uk} (GBD; corresponding author).}
Gavin B. Dalton\orcidlink{0000-0002-3031-2588},$^{2,3}$\footnotemark[1]         
Don Carlos Abrams,$^{1}$                                                        
Jure Skvarč,$^{1}$                                                              
Sergio Picó,$^{1}$                                                              
\newauthor%
Lilian Domínguez-Palmero\orcidlink{0009-0004-8470-9304},$^{1,4}$                
Illa R. Losada\orcidlink{0000-0002-0416-7516},\thanks{Independent Researcher;   
    WEAVE External Collaborator}
Sarah Hughes\orcidlink{0000-0002-7332-2751},$^{5}$                              
Neil O'Mahony,$^{1}$                                                            
Frank J. Gribbin,$^{1}$                                                         
\newauthor%
Andy Ridings,$^{1}$                                                             
David L. Terrett,$^{2,3}$                                                       
Cecilia Fariña\orcidlink{0000-0003-4940-3751},$^{1,4}$                          
Chris R. Benn,$^{1}$                                                            
Esperanza Carrasco\orcidlink{0000-0002-9174-5491},$^{6}$                   
\newauthor%
P. Joel Concepción Hernández,$^{1}$                                        
Kevin Dee,$^{1}$                                                           
Rafael Izazaga\orcidlink{0000-0002-3305-676X},$^{6}$                       
Shoko Jin\orcidlink{0000-0002-4824-8430},$^{7}$                            
Ian J. Lewis\orcidlink{0009-0009-2090-5513},$^{2}$                         
\newauthor%
J. Alfonso L. Aguerri\orcidlink{0000-0002-2839-2144},$^{4,8}$              
and
Gonzalo Páez\orcidlink{0000-0002-5609-6801}$^{9}$                          
\\
$^{1}$Isaac Newton Group of Telescopes, Apartado 321, 38700 Santa Cruz de La Palma, Tenerife, Spain\\
$^{2}$Oxford Astrophysics, University of Oxford, Keble Road, Oxford OX1 3RH, UK\\
$^{3}$RAL Space, Science and Technology Facilities Council, Rutherford Appleton Laboratory, Harwell Oxford, OX11 OQX, UK\\
$^{4}$Instituto de Astrof{\'i}sica de Canarias, Calle V{\'i}a L{\'a}ctea s/n, E-38205 La Laguna, Tenerife, Spain\\
$^{5}$Department of Physics and Kavli Institute for Astrophysics and Space Research, Massachusetts Institute of Technology, Cambridge, MA 02139, USA\\
$^{6}$Instituto Nacional de Astrof{\'i}sica, {\'O}ptica y Electr{\'o}nica, Luis Enrique Erro 1, C.P. 72840, Tonantzintla, Puebla, Mexico\\
$^{7}$Kapteyn Astronomical Institute, Rijksuniversiteit Groningen, Landleven 12, 9747 AD, Groningen, The Netherlands\\
$^{8}$Departamento de Astrof{\'i}sica, Universidad de La Laguna, Avenida Astrof{\'i}sico Francisco S{\'a}nchez s/n, E-38206 La Laguna, Spain\\
$^{9}$Centro de Investigaciones en {\'O}ptica, Loma del Bosque 115, Le{\'o}n, C.P. 37150, Guanajuato, Mexico
}

\date{Accepted 2026 March 31. Received 2026 March 11; in original form 2026 January 19}
\pubyear{2026}

\definecolor{bibref}{HTML}{B9529F}

\makeatletter
\newcommand{\@firstcitemark}[1]{%
  \ifcsname @cite@#1@used\endcsname
  \else
    \hypertarget{cite.first-#1}{}%
    \expandafter\let\csname @cite@#1@used\endcsname\@empty
  \fi
}
\patchcmd{\NAT@citex}
  {\@citea\NAT@hyper@{%
     \NAT@nmfmt{\NAT@nm}%
     \hyper@natlinkbreak{\NAT@aysep\NAT@spacechar}{\@citeb\@extra@b@citeb}%
     \NAT@date}}
  {\@firstcitemark{\@citeb}%
   \@citea\NAT@nmfmt{\NAT@nm}%
   \NAT@aysep\NAT@spacechar\NAT@hyper@{\NAT@date}}{}{}
\patchcmd{\NAT@citex}
  {\@citea\NAT@hyper@{%
     \NAT@nmfmt{\NAT@nm}%
     \hyper@natlinkbreak{\NAT@spacechar\NAT@@open\if*#1*\else#1\NAT@spacechar\fi}%
       {\@citeb\@extra@b@citeb}%
     \NAT@date}}
  {\@firstcitemark{\@citeb}%
   \@citea\NAT@nmfmt{\NAT@nm}%
   \NAT@spacechar\NAT@@open\if*#1*\else#1\NAT@spacechar\fi\NAT@hyper@{\NAT@date}}
  {}{}
\makeatother

\begin{document}

\label{firstpage}
\pagerange{\pageref{firstpage}--\pageref{LastPage}}
\maketitle

\begin{abstract}
We present the architecture, implementation, and on-sky validation of the fully
automated acquisition and guiding system (AG) developed for the WEAVE instrument
on the William Herschel Telescope.
The AG operates in two distinct modes, corresponding to the observing modes of
WEAVE. For the large integral field unit (LIFU), an off-axis imaging guider is
used, for which we have devised an automatic acquisition method based on pattern
recognition of stellar asterisms matched against Gaia predictions. For the
multi-object spectrograph (MOS) and the mini-integral field units (mIFU), a
multi-fibre guider uses up to eight coherent image guide fibre bundles to derive
and apply continuous corrections in azimuth, altitude, and rotation. The system
performs complete astrometric calculations, including atmospheric differential
refraction and instrument flexure, for each guide frame, enabling accurate
target placement and stable closed-loop guiding in all configurations. To
support development, commissioning, and operational validation, we have also
built a high-fidelity simulation mode that reproduces the behaviour of the
telescope control system and of the AG cameras, and we release the standalone
camera simulator as open-source software. Using two years of routine WEAVE
operations spanning commissioning and early survey phases, we present a
statistically robust characterization of AG performance, demonstrating that both
modes meet design requirements and are ready for sustained survey operations.
\end{abstract}



\section{Introduction}\label{sec:introduction}
WEAVE (WHT Enhanced Area Velocity Explorer) is a wide-field, multi-object,
fibre-fed, dual-arm survey spectrograph built for the Isaac Newton Group of
Telescopes' (ING) 4.2~m William Herschel Telescope (WHT) at the Observatorio del
Roque de los Muchachos (ORM) on the island of La Palma in the Canary Islands,
Spain (\citealp{dalton2012,dalton2016}). WEAVE was designed as a long-term
survey facility, enabling large spectroscopic programmes in Galactic
archaeology, stellar populations, galaxy evolution, and cosmology
(\citealp{jin2024}). Following integration and commissioning at the WHT during
the last few years (\citealp{dalton2020,dominguez2025}), WEAVE is now entering
survey operations, with stringent requirements on operational efficiency,
robustness, and repeatability. An important component in meeting these
requirements is the acquisition and guiding system (AG), which must reliably
position science targets on to fibres and maintain 0.3-arcsec stability over
long exposures across multiple observing modes. This paper describes the
architecture, implementation, and on-sky validation of the WEAVE AG system.

\begin{figure*}
  \centering
  \includegraphics[width=\textwidth]{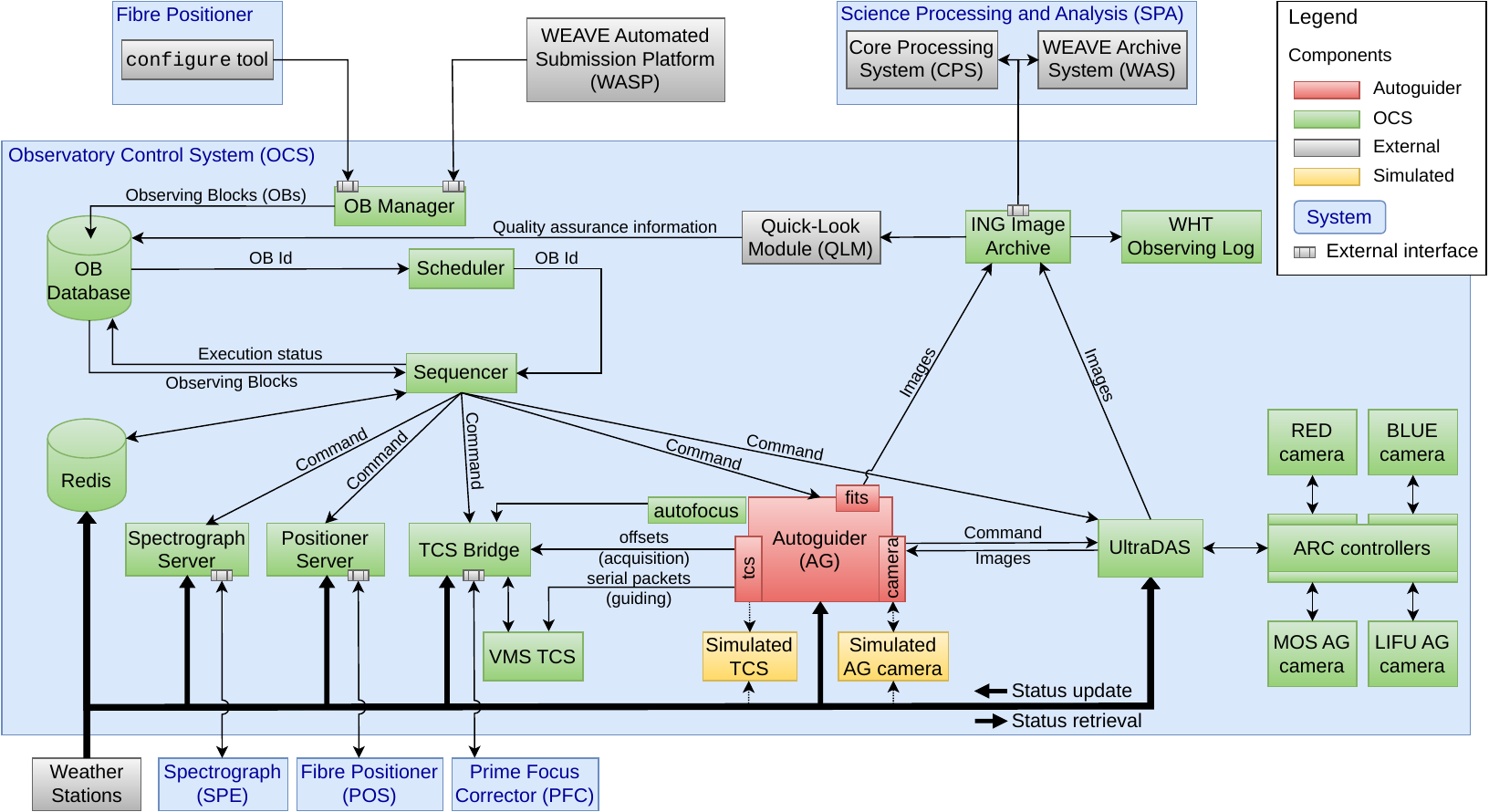}
  \caption{Top-level architecture of the WEAVE observatory control system (OCS)
  software and its interfaces to the wider WEAVE software environment. The AG
  subsystem, which is the focus of this paper, is highlighted in red. The
  diagram illustrates the principal communication pathways linking the AG system
  to other components, including the WEAVE sequencer, telescope control system
  (TCS), acquisition cameras, image archive, and Redis-based messaging
  infrastructure. These interfaces define the flow of commands, telemetry, and
  image data required to perform automated acquisition and closed-loop guiding
  during WEAVE observations. The diagram updates the architecture originally
  presented in fig.~1 of \citet{pico2018} to reflect the current system
  following integration and commissioning, and to emphasize the AG subsystem
  described in this work.}
  \label{fig:ocssoftware}
\end{figure*}

\subsection{The WEAVE design}\label{intro:weavedesign}
The design of WEAVE can be conceptually separated into four largely independent
subsystems: the prime focus corrector (PFC) with an integrated atmospheric
dispersion corrector (ADC), which delivers a 2\textdegree\ field of view (FOV)
(\citealp{agocs2012,agocs2014}); the rotator system used to compensate for the
field rotation during tracking (\citealp{delgado2018,sanvicente2020}); the
1000-multiplex fibre positioner (\citealp{lewis2014,hughes2023}); and the
dual-beam spectrograph, which provides a nominal resolving power of
$R \sim 5000$ over the full wavelength range of 366--959 nm in low-resolution
mode, and $R \sim 20\,000$ over a pair of restricted wavelengths in
high-resolution mode (\citealp{rogers2014}). The spectrograph has two cameras:
BLUE, optimized for wavelengths between 366--606 nm, and RED, for 579--959 nm
(\citealp{izazaga2018}). The entire WEAVE prime focus assembly can be translated
along the telescope optical axis and mechanically tilted by the focus
translation system (FTS), enabling both focus and tilt corrections to compensate
for the effects of temperature- and gravity-induced image degradation
(\citealp{canchado2016, tomas2018}).

WEAVE also prompted the creation of a new top-level observatory control system
(OCS) at the WHT (\citealp{pico2018}), enabling fully queue-based operations in
which observations are scheduled and executed automatically
(\citealp{farina2018}). \figref{fig:ocssoftware} presents a top-level
overview of the WEAVE OCS software architecture, highlighting in particular the
role and interfaces of the AG software within the OCS.
Several external subsystems interact with the OCS and therefore appear in the
architectural overview. Observations themselves originate upstream through the
WEAVE Automated Submission Platform (WASP), which handles the submission and
validation of survey targets against instrumental constraints and survey
requirements, before being ingested by the OCS as observing blocks (OBs). 
The Core Processing System (CPS) (\citealp{walton2014}) performs basic quality
control, image processing, and spectral extraction, while the WEAVE Archive
System (WAS) (\citealp{guerra2016}) provides long-term storage and controlled
user access to the data products -- with the OCS delivering its end-product (the
raw FITS files) to both CPS and WAS.
The Quick-Look Module (QLM) (\citealp{peralta2019}) operates alongside the OCS
during observations, providing rapid on-site assessment of data quality during
execution.
An important architectural element of the OCS is the Redis-based
noticeboard acting as a shared state layer between subsystems. By querying
the noticeboard rather than polling subsystems directly, the AG, sequencer,
and all of the other components have access to the real-time global system state
without introducing tight coupling between processes.

Central to the OCS is the concept of `sequences':
predefined procedures that orchestrate every step of an observation, from
telescope slews and instrument configurations to safety checks and data
acquisition, through the OCS sequencer software. The AG lies at the intersection
between the positioner (POS) and the OCS, providing the acquisition and guiding
corrections required for the sequencer to carry out the observations with
minimal human intervention.

\begin{figure*}
  \centering
  \includegraphics[width=\textwidth]{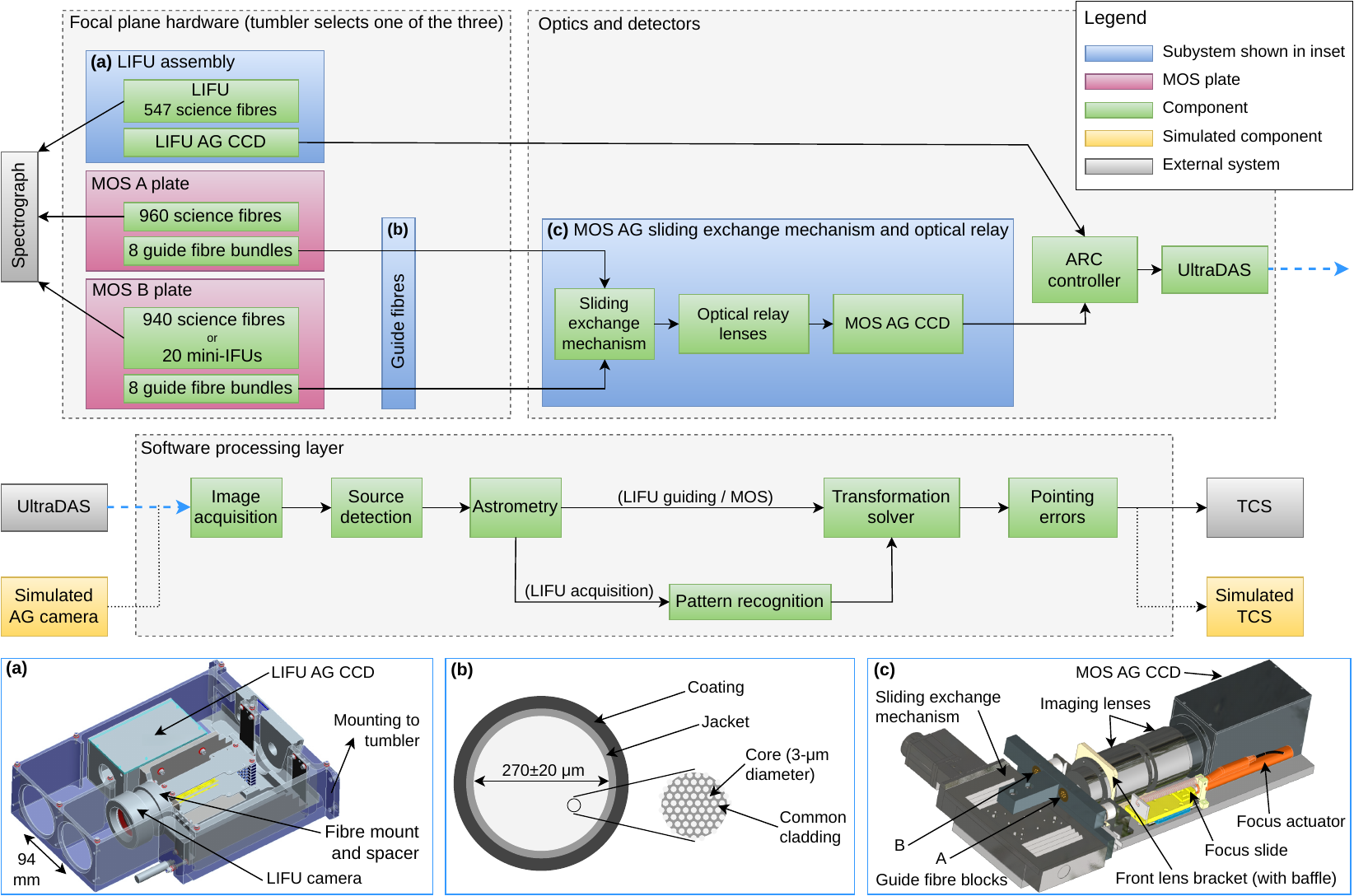}
  \caption{Top-level architecture of the WEAVE AG system. Light from the focal
  plane is captured by the LIFU AG CCD, or transmitted by the MOS guide fibre
  bundles, through an optical relay system, to the MOS AG CCD. The detectors are
  read out by an ARC controller that transmits the data to the acquisition
  software (UltraDAS). Images are then processed by the AG software to determine
  pointing and rotation offsets, which are sent to the TCS during acquisition
  and guiding. Both the camera and the TCS can be replaced with their simulated
  counterparts. Subsystems highlighted in blue are detailed in the insets:
  (a)~Computer-aided diagram (CAD) of the LIFU assembly, shown with the top
  plate removed; the fibre bundle is omitted for clarity. The distance between
  the centres of the LIFU AG CCD and of the LIFU (94 mm) is marked.
  (b)~Cross-section of a coherent MOS guide fibre bundle, redrawn from
  Fujikura FIGH-06-300PI specifications. (c)~CAD of the MOS AG sliding exchange
  mechanism, optical relay, and MOS AG CCD.}
  \label{fig:agdiagram}
\end{figure*}

At the hardware level, WEAVE has three interchangeable observing units at the
WHT primary focal plane (see \figref{fig:agdiagram}): two multi-object
spectroscopy (MOS) field plates and a large integral-field unit (LIFU). A
mechanical tumbler selects which of these three units is oriented toward the sky
during any given exposure (see fig.~6 of \citealp{dalton2016} for a schematic
representation of the tumbler assembly, illustrating the relative positions and
orientations of the two MOS field plates and the LIFU). The WEAVE fibre
positioner follows the buffered two-plate strategy pioneered by the 2dF
instrument (\citealp{lewis2002}), in which one MOS field plate is observing on
sky while the second is being configured for the subsequent observation
(\citealp{terrett2014,hughes2022a}). The two MOS plates, designated A and B,
provide 960 and 940 science fibres, respectively, with the reduced fibre count
on plate B reflecting the space required for the mini-IFUs (\citealp{jin2024}).
In addition to the science fibres, each plate also carries eight deployable
coherent guide fibre bundles (henceforth: `guide fibres') made of approximately
6000 cores used for acquisition and guiding, see inset (b) of
\figref{fig:agdiagram}. An example of a fully configured MOS field is presented
in fig.~2 of \citet{jin2024}, which illustrates a complete deployment of science
and guide fibres across the focal plane, while fig.~3 of \citet{hajnik2025}
shows the geometric and angular constraints governing the placement of guide
fibres. Configuring a full MOS field currently takes approximately 1~h and
45~min, while the design expectation for a fully optimized system is 1~h (with
the discrepancy arising primarily from multiple iterative adjustments required
to bring fibres within tolerance; work is ongoing to reduce the number of
iterations). The one-hour value sets both the typical duration of a WEAVE OB
being executed on the observing plate, and the time span over which the AG
system must maintain the required overall pointing stability (0.3-arcsec rms).
The relationship between the focal-plane hardware, the AG optics, and the
associated software control components is summarized in \figref{fig:agdiagram}.

In terms of observing modes, WEAVE supports three principal configurations
corresponding to different scientific use cases: MOS, mIFU, and LIFU. In MOS
mode, up to 960 or 940 (for MOS A and MOS B, respectively) independently
positioned fibres sample discrete targets across the full FOV for large
spectroscopic surveys of sparse source populations. In mIFU mode, deployable
mini-IFUs are positioned on selected targets in the FOV, providing spatially
resolved spectroscopy for up to 20 extended objects simultaneously, while
retaining wide-field multiplex capability. In contrast, LIFU mode uses a single
contiguous integral-field bundle comprised of 547 science fibres, centred on the
telescope pointing axis, to obtain spatially resolved spectroscopy of an
extended source -- or of multiple sources that fit in the LIFU FOV (e.g.,
\citealp{arnaudova2024}). Each observing mode places different requirements on
acquisition and guiding, including guide-star selection, focal-plane geometry,
and acquisition strategies, and these operational differences motivate the
design of the AG system described in the following sections.

The LIFU fibre head has a filling factor of $\approx 55$ per cent
(\citealp{jin2024}), leaving gaps between adjacent fibre cores. As a
consequence, a point source may fall entirely within a fibre, be only partially
intercepted, or fall between fibres and be effectively lost. To recover uniform
spatial coverage, LIFU observations are therefore typically performed using
dithering, in which successive exposures are taken with small telescope offsets
(of the order of 1 arcsec) following a hexagonal pattern. Combining multiple
dithered exposures ensures full sampling of the focal plane and is the standard
observing strategy for LIFU observations.

WEAVE also supports several non-standard modes.
In the case of MOS, WEAVE supports observations of low-target-density fields
using specialized WEAVE-Tumble-Less (WTL) OBs (\citealp{hajnik2025}).
In this mode, a single MOS field plate is pre-configured
to observe multiple independent sky fields within one observing block. Up to
three distinct science fields can be defined in advance, with the available
science fibres divided between them. A central guide fibre is shared between all
fields, while each field is associated with two additional, field-specific guide
stars. During the observing sequence the telescope executes blind offsets
between these predefined fields, successively placing each target region on the
configured fibre subsets without requiring plate reconfiguration. Guiding is
re-established at each position using the corresponding guide-star set. This
strategy enables efficient observations of low target-density surveys by
maximizing fibre utilisation and eliminating configuration overheads during
nighttime operations. This represents a distinct operational paradigm compared
to conventional MOS observations, in which the same science field and the same
set of guide stars are used throughout the entire exposure sequence.
In the case of LIFU, WEAVE supports non-sidereal observations of moving targets,
such as comets, asteroids, and interstellar objects. Special calculations during
guiding, together with differential tracking already supported by the WHT
telescope control system (TCS), allow WEAVE to maintain a non-sidereal target
at the centre of LIFU for as long as it is visible on sky (see
\secref{lifuag:nonsiderealguiding} below).

The large WEAVE FOV brings with it an additional challenge for the fibre
positioner and the AG: across 2\textdegree, the effects of atmospheric
differential refraction are significant (of the order of $\sim 1$ arcsec, larger
than the typical seeing at the WHT, even when the telescope is pointing close to
zenith; see \appref{app:diffrefr}), and must be taken into account. In addition,
for the expected typical exposure times with WEAVE ($\sim 1$~h), differential
refraction -- which is highly dependent on the zenith distance -- changes
continuously as the target follows its apparent motion on-sky. For this reason,
WEAVE observations are typically scheduled within a couple of hours of
culmination, and the fibre configuration is only valid for a comparable period
of time. Even so, the AG must continuously account for the change in
differential refraction and adjust the `guide pixel' inside each fibre as the
exposure progresses (see \secref{mosag:guiding} below).

During a MOS/mIFU exposure, the AG camera receives light from the appropriate
set of guide fibres via a sliding exchange mechanism, see inset (c) of
\figref{fig:agdiagram}. This brings the correct set in front of an optical relay
system consisting of two identical photographic lenses mounted front-to-front in
a perfectly symmetrical collimator/re-imager configuration, providing 1:1
magnification and transmitting the image from the output ends of the guide
fibres on to the AG CCD detector.

By contrast, in LIFU mode no guide fibres are deployed. Instead, a separate
guiding camera, adjacent and parfocal to the LIFU, looks directly at the sky
for acquisition and guiding, see inset (a) of \figref{fig:agdiagram}.

\subsection{The WEAVE AG system design}\label{intro:agdesign}
The WEAVE AG camera system consists of two $1072\times 977$ E2V 47-20 CCDs,
mounted in modified (for size, but otherwise standard) ING AG camera heads, with
a FOV of 3.8 arcmin and a pixel size of 13 ${\umu}\rm m$ (corresponding to 0.229
arcsec px$^{-1}$ at the WEAVE focal plane scale of 56.7 ${\umu}\rm m$
arcsec$^{-1}$). Due to constraints related to space, weight, and heat
dissipation at prime focus (and because it is not necessary to use both AG modes
simultaneously), both cameras are driven by a single Astronomical Research
Cameras (ARC) Generation III CCD controller (\citealp{leach2000}), 
as illustrated in \figref{fig:agdiagram}. The controller is mounted on the
exterior of the positioner structure on the top-end assembly,
connected via optical fibres to a Linux machine with a PCI-Express board, and
operated by a modified version of the in-house UltraDAS software
(\citealp{rixon2000}).
The images produced by UltraDAS are then passed to the AG software processing
layer (\figref{fig:agdiagram}), where they are analysed to determine pointing
errors that are subsequently transmitted to the TCS. Both the upstream
(UltraDAS) and the downstream (TCS) subsystems interfacing with this layer can
be replaced by simulated counterparts, enabling testing and development of the
guiding pipeline in a fully simulated environment (see \secref{sec:sim} below).

We stress that the AG itself does not apply any physical corrections to the
telescope. At the WHT, all guiding actuation is handled by the TCS, which
receives the measured pointing errors from the AG and converts them into small
positional offsets via its internal control loop. This loop is implemented using
a proportional--integral--derivative (PID) controller that continuously adjusts
the telescope's azimuth, elevation, and rotator axes to minimize the supplied
errors. During guiding, the AG therefore functions as a measurement and
error-reporting system: it determines the centroid shifts of the guide stars
on-sky and transmits these offsets to the TCS at each update cycle. The
separation of measurement (by the AG) and actuation (by the TCS) is a core
design feature of the WHT control architecture and ensures that the guiding
behaviour remains stable, predictable, and consistent with the telescope's
global control logic.

Because of its design, WEAVE requires two independent AG modes: the imaging LIFU
AG tool (\secref{sec:lifuag}) and the multi-fibre MOS/mIFU AG tool
(\secref{sec:mosag}), both incorporated into one application (the `WEAVE AG')
that performs all of the tasks. In order to test them well in advance of the
instrument commissioning, we devised a full simulation environment, including
simulated AG cameras and simulated TCS, which was extensively used during
development and integration testing, and is still being used for off-sky testing
of new features, improvements, and bug fixes (\secref{sec:sim}). From the start
of actual commissioning, the AG has seen constant on-sky usage and testing,
confirming both the accuracy and reliability anticipated from the simulation
mode and its compliance with the instrument specifications
(\secref{sec:results}), and providing the basis for the conclusions, lessons
learned, and future developments outlined in \secref{sec:discussion}.


\section{The LIFU Autoguider}\label{sec:lifuag}
The LIFU autoguiding mode (`LIFU AG') is a classical off-axis imaging autoguider
with a twist: its goal is to place the guide star on a specific pixel calculated
such that the central LIFU target falls on to a specific position (usually, but
not necessarily, the central LIFU fibre) during acquisition, and to keep the
guide star on the same pixel during guiding.

\begin{figure*}
  \centering
  \includegraphics[width=\textwidth]{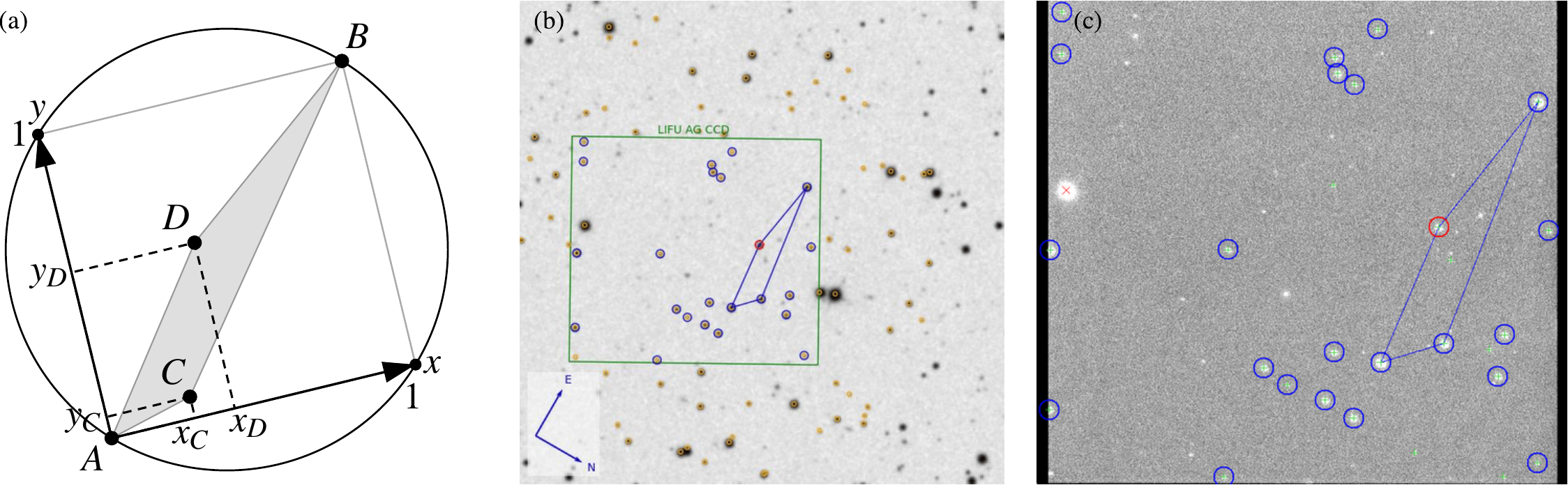}
  \caption{(a)~The geometric hash of a four-star asterism (shaded grey area)
  uses the most widely separated pair of vertices ($A,B$) to define a local
  coordinate system, and the remaining pair of vertices ($C,D$) to define a hash
  $\mathbfit{h}_{ABCD}$ (\eqrefp{eq:hABCD}) that is invariant under translation,
  scaling and rotation of the four stars.
  (b)~Typical finding chart presented by the LIFU AG; the background 
  is an 8-arcmin wide DSS2 image retrieved using \textsc{hips2fits}.
  The overlaid orange circles represent Gaia stars between which the reference
  asterisms are calculated during a successful LIFU acquisition, the finding
  chart also shows, with a solid blue line, the matching asterism validated by
  the stars overlaid by blue circles; the nominal LIFU guide star is overlaid by
  a red circle -- in this case it is one of the four stars forming the matching
  asterism, but that does not have to be (and often is not) the case; the green
  rectangle shows the extent of the LIFU AG CCD at the expected position (based
  on the desired telescope pointing calculated by the positioner software).
  (c)~Reduced image taken during the acquisition for WEAVE OB 8338 on 2023
  November 1. The box width is 4.1 arcmin (including the overscan regions).
  Overlaid on the image are solid blue lines outlining a confirmed asterism 
  -- compare with panel (b), blue circles around stars that validate the
  matching asterism, and a red circle around the nominal LIFU guide star.}
  \label{fig:hash}
\end{figure*}

\subsection{LIFU AG design}\label{lifuag:hardware}
The LIFU AG camera is mounted within the same physical structure as the LIFU
head, and they both rotate as a rigid body around the rotator axis (for a visual
reference, see inset (a) of \figref{fig:agdiagram}). There is a fixed distance
(94 mm, or $\simeq 23$ arcmin) between the centre of the AG camera ($y=-74$ mm
in plate coordinates) and the central LIFU fibre ($y=+20$ mm in plate
coordinates, see \appref{app:coords}). The focal plane is aligned to the
telescope pointing axis, which in practice is slightly offset (within a few
hundred $\umu$m) from both the plate centre and the rotator axis. A detailed
description of the rotator-axis alignment will be provided by G. Dalton et al.
(in preparation).

In principle, since the AG and the LIFU form a rigid system, if the exact
distance on-sky between the centres of LIFU and of the LIFU AG camera were known
with sufficient precision, the acquisition could involve offsetting the
telescope until the target fell precisely at the centre of the AG, and then
applying the (known) offset that would put the target at the centre of the LIFU;
one could then guide on any star visible in the AG FOV. However, such a process
is tedious and prone to introducing inaccuracies because of instrument flexure
and atmospheric differential refraction changing the on-sky distance between the
AG and the LIFU. In addition, it relies on the rotator angle matching the demand
value with high precision, and on the telescope offsets being executed with
correspondingly negligible translational error. The process would also fail if
the target were too faint to be seen on the AG detector, in which case blind
offsets would have to be used, introducing yet another source of errors. A new
acquisition system was therefore designed for LIFU.

\subsection{Automatic, pattern recognition-based acquisition}\label{lifuag:autoacq}
To acquire an object at the centre of the LIFU, a guide star must be selected
that, at the chosen sky position angle (PA), falls within the FOV of the AG
camera. Such a guide star is selected as part of the LIFU OB creation workflow,
and its coordinates are sent to the AG at the beginning of the OB execution.
The astrometric calculations that predict the position of the guide star on the
detector such that the target falls on the central LIFU fibre need to be
rigorous, because the acquisition is blind: one never sees the actual LIFU
target until after the spectra are taken. The matter is further complicated by
the LIFU having additional optics and hence a different plate scale than the
focal plane. These calculations are made by the positioner control software and
the AG in tandem: at every dithering step, the positioner calculates the
telescope pointing (internally called the `field' RA and Dec, or
\textsc{fldra} and \textsc{flddec}) such that the target falls at
the desired location on the LIFU, taking into account the additional optics
within LIFU and the instrument flexure; the AG, in turn, calculates the position
of the guide star such that the telescope indeed points to \textsc{fldra},
\textsc{flddec}, taking into account the instantaneous conditions such as
pressure, temperature, and relative humidity in order to account for the
differential refraction between the LIFU and the AG.

\subsubsection{Motivation}\label{lifuacq:motivation}
The large distance between the centres of the LIFU AG and of LIFU (94 mm) sets
strong requirements on the rotator precision: in order to have the target within
0.3~arcsec (or 19.5 ${\umu}\rm m$ at the LIFU plate scale) of the desired
position on the LIFU, the error in rotation must not exceed
$\sim 0.01$\textdegree\ (assuming no errors in $x$ and $y$ in the tangent
plane), which we found requires corrections in rotation during the acquisition
process. More than one star must therefore be used for the LIFU acquisition,
which represents a departure from the classical IFU single-star acquisition
paradigm.

This is implemented by the AG automatically detecting stars in the FOV,
determining where they should actually appear on the detector, based on the
desired telescope pointing and the full astrometric calculations, and applying
corrections to all three telescope axes iteratively, until the errors fall below
a specified threshold.

Once the LIFU acquisition starts, the software performs the following background
operations while the camera integrates and reads out the first acquisition
image (cf. the software processing layer of \figref{fig:agdiagram}):
(1) it retrieves the current time, telescope position, and latest weather
measurements, precomputes certain quantities necessary for the astrometry, and
performs the astrometry for the guide star as described in \appref{app:coords},
determining its expected plate coordinates under the assumption that the focal
axis of the telescope points exactly where instructed to by the positioner;
(2) it retrieves the finding chart image from \textsc{hips2fits}, and the list
of stars expected in the finding chart from the Gaia database, in parallel; the
exact surveys (typically DSS2 Red for the finding chart image and Gaia DR3 for
the list of stars), FOV, magnitude limits, maximum number of stars, etc. are all
configurable;
(3) once the list of stars is retrieved, it performs the full astrometric
calculations for each of them, determining the expected plate coordinates of
each star in the finding chart;
(4) it then applies the algorithm described below to precompute hashes of
four-star asterisms in plate coordinates.

\subsubsection{Pattern recognition algorithm}\label{lifuacq:langetal}
\citet{lang2010} introduced a robust algorithm for blind astrometric calibration
of arbitrary astronomical images: given an image with no information apart from
the pixels themselves (not even the image scale), the algorithm can reliably
determine the pointing, scale, and orientation of the image.

The algorithm works by comparing `hashes' of four-star asterisms detected
in the image with precomputed asterism hashes from catalogue data,
see \figref{fig:hash}. Given an asterism $ACBD$ with four stars as vertices, its
hash is computed as follows: the most widely separated
pair of vertices ($A$ and $B$) define a local coordinate system where $A$ is at
$(0,0)$ and $B$ is at $(1,1)$, and in which
the four transformed coordinates of the remaining two vertices
form a geometric hash,
\begin{equation}\label{eq:hABCD}
  {\mathbfit h}_{ABCD} = \begin{bmatrix}x_C & y_C & x_D & y_D\end{bmatrix},
\end{equation}
\noindent that is invariant under translation, scaling, and rotation of the four
stars. In order for the hash to be unambiguous with respect to swapping $A$/$B$
and $C$/$D$, the following two additional constraints are imposed when choosing
the labels: $x_C \leq x_D$ and $x_C + x_D \leq 1$.

The invariance of the hashes to conformal transformations makes this an ideal
pattern recognition algorithm for automatic acquisition. The `reference' hashes
are calculated from the expected positions of Gaia stars within the FOV of the
camera, following the full astrometric calculations, from mean International
Celestial Reference System (ICRS) (\citealp{arias1995}) to plate
coordinates. Meanwhile, the `measured' hashes are calculated from the measured
centroids in the acquisition image, converted from pixel to plate coordinates.
Reference and measured hashes are then compared until two similar hashes (within
a configurable tolerance margin) are found, and the rigid transformation between
them (\appref{app:transrot}) yields the required correction in $x$ and $y$ (in
the focal plane) and rotator angle.

The original algorithm was used to hash all-sky (or near-all-sky) optical
surveys containing over a billion stars, such as USNO-B1 (\citealp{monet2003}),
in order to process input images representing an arbitrary patch of sky at an
arbitrary scale. The requirements for the LIFU acquisition tool are much more
restrictive, thus simplifying the algorithm: the FOV is typically known to
within several arcsec, while the scale is constrained by the size and pixel
scale of the detector. The Gaia stars (typically tens or hundreds) expected to
appear in the acquisition image can therefore be retrieved, and the four-star
asterism hashes can be precomputed on-the-fly, during the first acquisition
exposure.

Similar pattern-recognition approaches have previously been used for blind field
acquisition in robotic telescopes (e.g.,
\citealp{watson2012,bakos2013,schindler2016,reinacher2018,husser2022,ahn2024,galliher2024,gill2024}).
In sparse fields, the small number of available guide stars can render direct
application unreliable, motivating the development of customized acquisition
algorithms adapted to specific operational constraints (e.g.,
\citealp{foale2018,rosenberg2018,price2024}). The LIFU implementation shares
this constraint-driven approach by operating on a small, off-axis guide field
and performing the matching directly in plate coordinates, using real-time
Gaia-based asterism generation.

\subsubsection{Image processing}\label{lifuacq:imageproc}
Once the first acquisition image is read out by the camera, the software detects
the centroids of all stellar-like sources in the image using \textsc{photutils}
(\citealp{bradley2025}), and converts them to plate coordinates
(\appref{app:coords}). It then proceeds to compare every measured four-centroid
asterism (${\mathbfit h}_{\rm m}$) against the precomputed reference hashes
(${\mathbfit h}_{\rm r}$),
\begin{align}
  {\mathbfit h}_m &= \begin{bmatrix}x_{C,{\rm m}} & y_{C,{\rm m}} & x_{D,{\rm m}} & y_{D,{\rm m}}\end{bmatrix}\\
  {\mathbfit h}_r &= \begin{bmatrix}x_{C,{\rm r}} & y_{C,{\rm r}} & x_{D,{\rm r}} & y_{D,{\rm r}}\end{bmatrix},
\end{align}
\noindent by calculating the absolute error of each of the four numbers in the
hash,
\begin{align}
  \varepsilon_{x,C} &= |x_{C,{\rm m}}-x_{C,{\rm r}}|\\
  \varepsilon_{y,C} &= |y_{C,{\rm m}}-y_{C,{\rm r}}|,
\end{align}
\noindent etc., and the total rms error of the hash,
\begin{equation}
  \varepsilon_{\rm rms} = \left(\varepsilon_{x,C}^2 + \varepsilon_{y,C}^2 +
                                \varepsilon_{x,D}^2 + \varepsilon_{y,D}^2\right)^{1/2}.
\end{equation}
\noindent Every asterism for which all four errors $\varepsilon_{x,C}$, etc., as
well as the total $\varepsilon_{\rm rms}$, are below a configurable threshold is
flagged as a match, and the software computes its associated offset (in
translation and rotation) by performing a rigid transformation between the two
sets of points (\appref{app:transrot}). If this is successful, the
transformation $(t_x, t_y, \phi)$ is applied to all other measured centroids,
which in principle should yield the positions of reference stars. If several
stars are `recovered' in this way (in other words, if the software can determine
with certainty which reference stars correspond to which measured centroids),
the asterism is deemed confirmed; \figref{fig:hash} shows how all such confirmed
stars are marked with a blue circle, and it is easy to verify by eye that the
pattern of stars is the same in both catalogue data (b) and in the detector
image (c). When a transformation is confirmed, it is applied to the telescope as
an offset in the tangent plane, plus an offset in rotator angle, and the
operation is repeated until the calculated telescope offsets fall below a
specified threshold.

\subsubsection{Failure modes and mitigation strategies}\label{lifuacq:failure}
The algorithm can fail for a number of reasons:

\begin{description}
\item \textit{Sparse field:} if there are fewer than four detectable stars in
the FOV, no asterism can be computed; this situation is foreseeable (since the
number of stars in the FOV can be predicted based on Gaia data), and the
algorithm will proactively determine whether there exists a small telescope
offset (up to a few arcmin) that will bring more stars into the FOV without
moving the guide star too far from its desired position; in practice this always
solves the problem, since it is highly unlikely to have such a sparse
acquisition field so as not to find more stars in any direction.
\item \textit{Crowded field:} if there are too many precomputed hashes and too
many detected centroids (e.g., in an extremely crowded acquisition field), a
naive application of the algorithm can be prohibitively computationally
expensive (the time limit for finding a matching asterism in a single
acquisition frame is configurable, but a typical value is 10 s).
\item \textit{Clouds:} if enough stars are expected in the FOV, but they are not
detected, this is likely due to thin clouds, or simply because the exposure time
is too short; the algorithm will automatically increase the exposure time and
reattempt to acquire, which normally solves the problem; otherwise, the OB can
be repeated at a later time.
\end{description}

In order to avoid such situations, certain mitigation measures are in place:
(1) hashes are only precomputed for four-star asterisms within a reasonable
range of scales (e.g., the separation between any two stars should not be
smaller than several pixels, nor larger than the longest side of the detector);
(2) both the catalogue stars and the centroids are sorted heuristically based on
a merit function that combines magnitude and distance to the centre of the
detector, such that high-signal-to-noise (SNR) stars likely to be in the FOV
are checked first;
(3) in very crowded fields, the range of magnitudes for the stars retrieved from
Gaia and used to calculate the reference hashes is automatically reduced, in
order to lower the number of asterisms and obtain a match within a reasonable
time;
(4) centroids that may turn out to be problematic in the input image (e.g.,
overexposed stars, stars that are very close together, stars that fall on a
known bad column, etc.) are ignored unless the field is extremely sparse.

These measures have been developed over the course of LIFU commissioning, during
investigations of the algorithm's failure modes. In the rare cases that
automatic acquisition still fails, the OB sequence must be aborted by the user;
we found that the lack of corrections in rotation can potentially result in
pointing errors large enough that the precision required by the LIFU
specifications (0.3-arcsec rms) is no longer attained.

A particular challenge, inherent to off-axis guiders (and shared by both LIFU
acquisition and guiding) is the difficulty to measure field rotation with high
precision, because the available guide stars occupy a compact region that is
poorly distributed about the rotator axis. In the case of LIFU AG, the guide
camera samples a small patch of the focal plane displaced by 74 mm in $y$ from
the plate centre, so the stellar centroids lie within a narrow cluster. For such
a geometry, a small field rotation produces nearly identical displacements for
all stars, which are therefore highly degenerate with pure translations in the
$x$ direction, rendering the rigid-body transformation intrinsically
ill-conditioned. Even small stochastic centroid variations (dominated
by atmospheric seeing fluctuations, photon noise, and detector noise) can map
directly into relatively large, rapidly varying rotation estimates because of
the intrinsic translation--rotation degeneracy. As a consequence, the rotation
term becomes dominated by estimator noise rather than true mechanical rotation
of the telescope, leading to unstable or oscillatory corrections if naively fed
back to the mount. This geometric degeneracy explains why acquisition and
guiding in rotation is fundamentally more difficult for off-axis systems (such
as the LIFU AG) than for guiders that sample multiple, widely separated field
points (such as the MOS AG).

Although the same geometric degeneracy affects both LIFU acquisition and
guiding, the practical consequences differ significantly. During acquisition,
the algorithm is executed in a small number of discrete steps to obtain a
single, best-fit solution that places the science target at the centre of LIFU.
In this regime, the inherent translation--rotation coupling is typically
tolerable: even if individual solutions are affected by noise, the iterative
acquisition procedure applies stringent convergence thresholds on the solution,
so additional acquisition steps are simply performed until all errors fall below
their limits and accurate target placement is achieved. During closed-loop
guiding, however, the situation is more problematic. Because the solution is
continuously recomputed and fed back to the telescope, noise-driven fluctuations
in the poorly constrained rotation term propagate directly into the applied
corrections and can produce persistent oscillations about the nominally correct
position. These oscillations, often at the level of several tenths of an arcsec,
do not represent gross mis-pointing but instead cause effective image motion
during the exposure, leading to a measurable degradation of image quality and
point spread function (PSF) broadening. For this reason, although rotational
corrections are applied successfully during acquisition, their use in
long-duration guiding remains under active investigation.

Some possible approaches during guiding are:
(1) strongly damping or suppressing rotation in the control loop;
(2) temporal filtering of the rotation solution, using low-pass or Kalman
filtering (\citealp{kalman1960}), so that only long time-scale, coherent
rotation signals are passed to the mount, while frame-to-frame noise is
rejected;
(3) monitoring the conditioning of the plate-solution matrix in real time, such
that when the fit becomes degenerate the AG can automatically dampen or disable
rotation updates for that frame.

Another potential source of acquisition difficulty arises 
in the case of close double stars in the Gaia catalogue, which may appear
partially blended or unresolved on the detector, introducing small systematic
centroid offsets. Partially resolved pairs are deblended and generally rejected
by the minimum-separation filtering applied to detected sources, while
unresolved pairs behave as single centroids within the effective AG camera
resolution. If centroid measurements nevertheless become inconsistent with the
expected pattern solution, the acquisition algorithm converges on an alternative
asterism. This effect therefore represents a secondary limitation compared to
the other constraints discussed above.

Finally, a further potential concern is whether large colour differences between
stars within a guiding asterism could introduce apparent geometric distortions
through atmospheric differential refraction (\appref{app:diffrefr}). However,
the WEAVE PFC is equipped with an ADC that compensates atmospheric dispersion
over 370--1000 nm up to zenith distances of 65\textdegree\ 
(\citealp{agocs2014}). Since both MOS and LIFU acquisition and guiding operate
downstream of the PFC+ADC, first-order colour-dependent image shifts are
removed prior to detection on the AG camera. Residual chromatic differential
refraction after ADC correction introduces relative centroid shifts below
$\approx 0.1$ arcsec (less than half a pixel) over the full zenith-distance
range, not enough to measurably distort the asterisms.

\subsection{Guiding}\label{lifuag:guiding}
Similar to the LIFU acquisition (\secref{lifuag:autoacq}), the ultimate goal of
LIFU guiding is to ensure that the optical axis of the telescope keeps pointing
at the field coordinates \textsc{fldra} and \textsc{flddec} calculated by the
positioner such that the LIFU target falls on to the chosen LIFU fibre. Dithered
LIFU observations are performed by adjusting the \textsc{fldra} and
\textsc{flddec} accordingly, with no modification to the AG procedures.

WEAVE OBs are typically divided into multiple exposures of at most 20 minutes
each, which may or may not be dithered. In between these, the OCS always
suspends guiding, dithers the telescope (if applicable), and then resumes
guiding. Upon resumption, the WEAVE AG recomputes the guide pixel, taking into
account not only any possible dithering offset, but also a new estimate of the
differential refraction. The contribution of differential refraction on the LIFU
guide pixel is calculated at the beginning of each science exposure, and is then
kept constant throughout it. This is in contrast to the MOS mode
(\secref{mosag:guiding}), where the guide pixel is recomputed for every single
autoguiding frame, thus differential refraction is properly accounted for
throughout the entire OB. While this algorithm has been proven to fulfil the
guiding accuracy required by the instrument specification (0.3-arcsec rms), a
planned future improvement in LIFU guiding is the continuous update of the guide
pixel to reflect instantaneous changes in the differential refraction.

\begin{figure}
  \centering
  \includegraphics[width=\columnwidth]{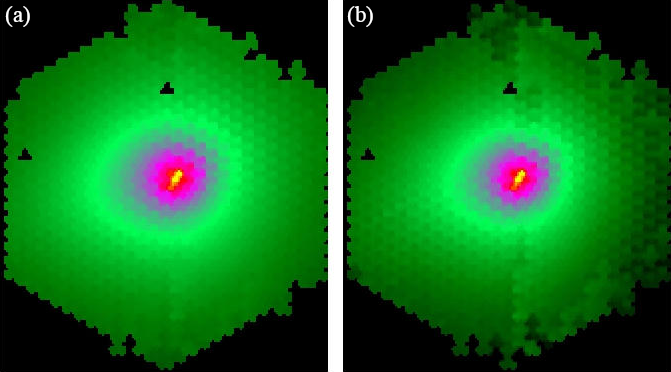}
  \caption{Collapsed and sky-subtracted images of comet 3I/ATLAS observed on
  2025 November 29 with LIFU in low-resolution mode. Panels (a) and (b) show
  the images obtained with the BLUE and RED cameras, respectively.}
  \label{fig:comet}
\end{figure}

The WEAVE requirements only demanded the application of LIFU guiding corrections
in $x$ and $y$, in order to keep the guide star on the guide pixel, without
performing any correction in rotation. The primary reason was that the WEAVE
rotator was designed to track accurately enough over the course of a typical
LIFU exposure ($\sim 20$ minutes). In addition, differential refraction does not
produce a noticeable skewing over the given time-scales and across the FOV of
the AG camera (less than 4 arcmin), in contrast to the guide fibres in MOS,
which are typically distributed across the 2\textdegree\ FOV. Nevertheless,
since corrections in rotation are already applied during LIFU acquisition, we
are currently implementing them into the guiding algorithm as well, subject to
the constraints described above.

\subsection{Non-sidereal acquisition and guiding}\label{lifuag:nonsiderealguiding}

The introduction of a non-sidereal guiding mode for LIFU observations was
motivated by the increasing demand for differential tracking of moving targets,
such as comets, asteroids, and interstellar objects. WEAVE observations assume
sidereal tracking, where the telescope follows the apparent motion of fixed
celestial sources. However, during the 2025 summer campaign targeting comet
3I/ATLAS (\citealp{denneau2025}), a series of manual observations demonstrated
that high-quality LIFU data could be obtained with differential tracking, albeit
through labour-intensive and error-prone procedures. The success of these
observations highlighted the scientific potential of systematic non-sidereal
campaigns, while also exposing the need for a dedicated, automated OB sequence
capable of performing such observations with minimal observer intervention. This
has recently been implemented, and used to successfully acquire 3I/ATLAS at the
centre of LIFU (\figref{fig:comet}) as a proof-of-concept.

\begin{figure*}
  \centering
  \includegraphics[width=\textwidth]{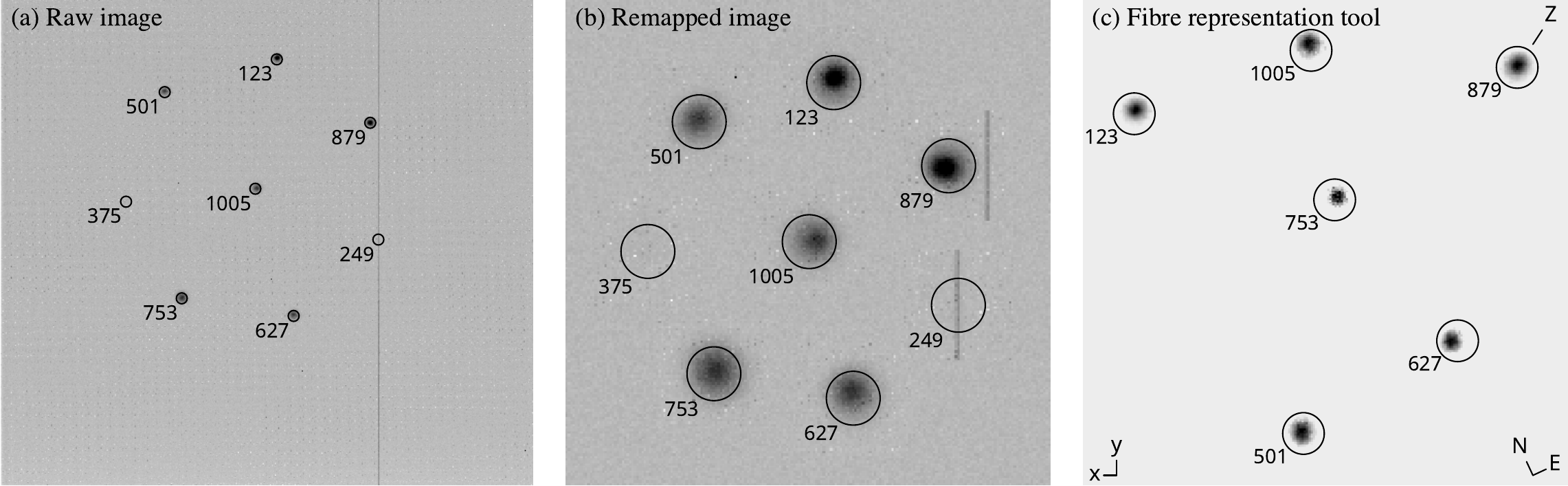}
  \caption{(a) Raw image taken during the acquisition for WEAVE OB 21780 on 2025
  December 2.
  (b) The same guide fibre data, remapped in their original relative positions
  on to a smaller raster (by cropping the fibre images and pasting them close
  together, with the empty space filled with artificial noise with the same mean
  and standard deviation as the bias of the real camera), useful for real-time
  display in the control room.
  (c) Fibre representation tool, with the fibre images processed as described in
  the main text (\secref{mosag:fibrerepr}).
  In all three panels, the fibre outlines and the fibre positioner IDs are
  overlaid on top of the image. Also, the original colours (white on black) are
  inverted to improve visual clarity.}
  \label{fig:mosccd}
\end{figure*}

The goal of the non-sidereal mode is to allow the LIFU AG to track moving
targets automatically, by applying differential rates in right ascension and
declination derived from target ephemerides. This capability enables long
integrations on fast-moving objects without smearing, ensuring that the science
target remains in the central LIFU fibre throughout the exposure. The
implementation is designed around an enhanced OB sequence that incorporates the
required time-dependent parameters (RA, Dec, and differential rates) directly in
the OB definition. The OCS coordinates the execution to acquire the field in
advance, activates the TCS differential tracking at the scheduled observing
time, and then continues guiding while accounting for the relative motion of the
guide star in the AG frame.

From a software architecture perspective, the guiding logic remains consistent
with the existing WEAVE acquisition and guiding framework, minimizing disruption
to the broader control system. The WEAVE AG's core principle (of maintaining the
telescope pointing at \textsc{fldra} and \textsc{flddec}, see
\secreftwo{lifuag:guiding}{mosag:guiding}) is perfectly suitable for
continuously changing sky positions. Instead of holding \textsc{fldra} and
\textsc{flddec} constant during a guiding cycle, the system can update them
dynamically, allowing the telescope to follow the target's apparent motion
across the sky. This approach preserves the existing interface between the AG
and the positioner subsystems, ensuring compatibility with dithering and normal
AG operations. In practice, an auxiliary script updates \textsc{fldra} and
\textsc{flddec} in real time based on the target's ephemerides and current
Modified Julian Date (MJD),
including the necessary geometric transformations accounting for LIFU's
different plate scale and optical distortions.

The implementation of non-sidereal guiding for LIFU broadens the range of
observing modes supported by WEAVE, allowing automated observations of Solar
System and interstellar targets in time-critical programmes.


\section{The MOS/mIFU Autoguider}\label{sec:mosag}
The MOS/mIFU autoguiding mode (`MOS AG') uses images from the guide fibres
placed by the positioner on the field plate while preparing the MOS (A/B) or
mIFU observations; the goal of this software is to place the stars at specific
positions inside each guide fibre by applying telescope offsets during
acquisition, and to keep them in specific positions during guiding. We refer to
`specific positions' instead of `centres' because the AG must also account for
differential refraction and for dithering. Each guide star travels along a short
trajectory inside its respective guide fibre, which depends on its change in
altitude during the science exposure. Due to the large FOV, the differential 
refraction between the guide fibres is not negligible (see
\appref{app:diffrefr}) and must be accounted for. 

\subsection{The fibre representation tool}\label{mosag:fibrerepr}
The diameter of the guide fibres is 270 ${\umu}\rm m$ (or $\sim 4.8$~arcsec on
sky), as shown in inset (b) of \figref{fig:agdiagram}. Given the detector pixel
scale of 13 ${\umu}\rm m$~px$^{-1}$, this translates into $\sim 21$~px, or about
2 per cent of the full extent of the detector. The separation between the guide
fibres is constrained by the ferrules that encase their end sections, with a
diameter of 2.5 mm ($\sim 192$~px), which is also the minimum distance
physically possible between any two guide fibres (although in practice there is
some tolerance on top of that, so that the typical distance between two adjacent
guide fibres on the detector is $\sim 230$~px).

Any unprocessed image of the guide fibres on the detector, even if tightly
cropped, will therefore show the guide fibre images far too small for the user
to distinguish any features, see panel (a) of \figref{fig:mosccd}. If the AG
were to also overplot the necessary additional markings (e.g., the measured
centroid and the expected position of the star), the image would then become
impossible to decipher.

In order to improve the user experience, the AG crops a portion of the detector
image surrounding each guide fibre, and remaps the pixels on to a smaller
raster, see panel (b) of \figref{fig:mosccd}; we refer to this as the `remapped
image`, and it is the one actually displayed in \textsc{ds9}
(\citealp{joye2003}) on the control screen. This procedure magnifies the fibre
images, so that small motions of the guide stars can be easily recognized and
the overlaid markings can be read. The remapped image is effectively a faithful
representation of the raw image, but without the empty space between and around
the fibres.

In addition, the AG also creates a third image (displayed directly in the AG
window), shown in panel (c) of \figref{fig:mosccd}, which we refer to as the
`fibre representation tool'. The same cropped fibre images used to generate the
remapped view are further processed and combined in order to:
(1) normalize the signal in each fibre so that all stars have the same peak
intensity; 
(2) de-rotate each fibre's raster to compensate for the fibre orientations (see
later \secref{calibs:fibreorient}), ensuring that North is aligned consistently
across all fibres and that the guide stars behave predictably when the telescope
is offset; and 
(3) reposition the fibre images so that their relative locations on the plate
are correctly reproduced, with the fibre closest to the plate centre appearing
closest to the centre of the AG image. At the corners of the image, the AG marks
the plate coordinate axes ($x$ and $y$; bottom--left), the direction of North
and East (bottom--right), as well as the direction of Zenith (top--right). The
rotation and flip are always calculated so that $x$ and $y$ point as shown in
the figure, since this is the typical display convention used by the positioner
software.

Unlike the raw or remapped CCD images, which always show all eight guide fibres,
the fibre representation tool displays only those guide fibres that are
configured and deployed on the plate for the current observation, since parked
fibres do not have meaningful plate positions. For example, in
\figref{fig:mosccd}, although all eight guide fibres are marked in panels (a)
and (b), two are empty; this is immediately apparent in the fibre representation
tool in panel (c), where only the six configured fibres are displayed. The fibre
representation is therefore a faithful representation of the guide stars on the
plate, and we found it operationally useful because it provides an immediate and
intuitive view of the global guiding geometry, allowing users to recognise
coherent motions, identify anomalous guide fibres at a glance, and diagnose
acquisition or guiding issues that would be difficult to interpret from the
individual fibre views alone.

\subsection{Automatic acquisition}\label{mosag:autoacq}
Every MOS AG cycle, whether during acquisition or guiding, begins with an image
being taken by the MOS AG camera (see the software processing layer of
\figref{fig:agdiagram}). The positions and orientations of the guide
fibres on the CCD are determined in advance as part of the fibre calibration
procedure (\secref{mosag:calibs}). Once the image is read out, the software
subtracts the bias (estimated as the sigma-clipped median of the pixels outside
the guide fibres) and crops out the image produced by each fibre (represented by
a black circle in \figref{fig:mosccd}); it then proceeds to estimate the sky
level inside each fibre, by determining the centroids, masking the stars, taking
the sigma-clipped median of the remaining pixels, and subtracting it from the
fibre image. Both the bias and the sky-background standard deviations are stored
in order to produce later on an estimated SNR for each guide star.

Once the centroids are known, they are converted to plate coordinates
(\appref{app:coords}) using the measured offsets from the fibre centres on the
CCD, the fibre orientations, and the known position of each fibre on the plate;
these are the `measured' plate coordinates of the stars. The `expected' plate
coordinates are determined by performing the full astrometric calculations, from
mean ICRS coordinates to plate coordinates, taking into account the current
time, differential refraction (based on the current weather conditions; see
\appref{app:diffrefr}), and any potential offsets in the case of dithered
observations.

Although the centroiding procedure is robust, on-sky tests have shown that, in a
small fraction of cases, guide fibres may be positioned suboptimally on the
plate. Misplacements of up to $\sim 1$~arcsec have been observed on rare
occasions, sufficient for the guide star to fall noticeably off-centre within
the fibre. In such cases the resulting stellar profile becomes asymmetric,
yielding degraded or biased centroid estimates even when the SNR is high. The
user has the choice of excluding these guide fibres during acquisition and
guiding, and let the solution be driven by the remaining, well-centred stars. 
These events, although infrequent and currently under active investigation,
represent a mechanical rather than algorithmic limitation, and explain
occasional cases where the MOS AG does not reach the 0.3-arcsec rms accuracy
imposed by the specifications. In practice, such instances manifest as small but
measurable cases of suboptimal fibre positioning. These arise either when a
fibre placement does not fully converge, leaving the fibre slightly offset, or
when a fibre positioned at a specific angle is subsequently disturbed by the
release of a neighbouring fibre. Although rare, these effects affect the current
performance limitation of the system and motivate the follow-up work described
in \secref{discussion:future}.

The required telescope guiding corrections are calculated from the measured and
expected guide star positions, according to the rigid-transformation algorithm
described in \appref{app:transrot}. In order to successfully calculate the
rotation offset, the algorithm needs at least two points; if only one centroid
is detected, only the translation offset required to place that one centroid on
to its expected position is sent to the TCS, and no correction in rotation is
made.

Sometimes, just after pointing the telescope to a new field, no stars are
visible in the guide fibres, but usually some light from a star falls into at
least one fibre; if there is just enough light to meet the SNR threshold (which
can be lowered to as little as 2), the AG will `pull in' that star a bit closer
to the fibre with every acquisition cycle; once light from a second star is
visible, the algorithm can start applying corrections in rotation, and will then
pull in the rest of the stars and converge extremely fast, usually within 1--2
additional steps.

If, however, after pointing at a field, there is no signal whatsoever in any of
the guide fibres, the automatic acquisition fails immediately with a warning,
and the user can either deploy the focal plane imager and manually put one or
two of the guide stars inside their respective guide fibres, or perform a blind
spiral search.

\subsection{Fallback to manual acquisition using the FPI camera}\label{mosag:fpiacq}
The focal plane imager (FPI) system consists of a sky-viewing camera and a
fibre-viewing camera, mounted on an $x$--$y$ gantry operating between the last
element of the prime focus corrector and the focal plane (\citealp{hughes2020}).
Both cameras can access the entire focal plane, but the FPI blocks part of it,
so other operations (acquisition, guiding, observations) are limited or even
impossible while it is deployed. The sky-viewing camera, with a FOV of
approximately $2.4\times 2$ arcmin, was primarily designed to perform
astrometric calibrations of the focal plane, but its secondary uses include
(backup) field acquisition and aiding with standard star observations. The FPI
design was inherited from and closely follows that of the 2dF focal plane imager
(\citealp{lewis2002}).

Due to the small size of the guide fibres ($\sim 4.8$~arcsec across) in relation
to the large FOV of WEAVE ($\sim 2$\textdegree\ across), even small pointing (or
astrometry) errors in rotation (of the order of $0.03$\textdegree) are
sufficient to place the guide stars outside of their respective fibres, making
it impossible to acquire with the AG alone. When that happens, the FPI can be
deployed at the known position of a guide fibre on the plate. An image is then
taken, and the guide star will normally be seen at some distance (typically
several arcsec) from the desired position. The telescope can then be offset to
cancel out this error. Typically, two stars are used for the FPI: one near the
centre of the plate for estimating the $x$ and $y$ offsets, and another star
further out near the edge of the plate for estimating the rotation offset. With
at least two guide stars now guaranteed to appear inside their respective
fibres, the FPI can be retracted and the automatic acquisition can resume. For
an experienced user, the entire procedure of acquiring two guide stars with the
FPI generally takes under three minutes.

To support manual acquisition with the FPI, the OCS provides a dedicated finding
chart tool. When activated, it displays up to eight small Aladin-based charts,
each showing the FPI FOV as it would appear when centred on one of the deployed
guide fibre positions. The guide stars are marked, allowing the observer to
identify unambiguously which star should be selected and placed on each guide
fibre. This functionality is particularly useful in crowded fields, where
multiple stars may fall within the FPI FOV.

\subsection{Fallback to spiral search}\label{mosag:spiralacq}
A spiral search involves automatically offsetting the telescope in predetermined
increments (e.g., few arcsec) that trace a coarse spiral around the starting
point, while taking one exposure at each position; once the faintest signal is
detected in even one of the guide fibres, the automatic acquisition takes over
and pulls the remaining stars inside their respective fibres, as described
before. This is a last-resort option, since it is always much faster to deploy
the FPI and manually acquire one or two stars as described above, than to
blindly search around the initial position: the number of spiral search
positions increases as the square of the distance, and half of these movements
are in the wrong direction. Nevertheless, the spiral search can be used with
some success if the steps are significantly larger than the diameter of the
guide fibre ($\sim 4.8$~arcsec), since the AG is sensitive enough to pull in
stars that are outside of (but close to) their respective fibres.

\subsection{Guiding}\label{mosag:guiding}
Once the MOS acquisition has been deemed successful based on the total
corrections (in $x$, $y$, and rotation) and the rms error all falling below
predefined, user-configurable thresholds, the AG can proceed with closing the
guide loop. For MOS, the algorithms for automatic acquisition and for guiding
are identical, so the procedure detailed in \secref{mosag:autoacq} is merely
repeated until guiding is stopped. The differences are mostly bookkeeping. The
AG records guiding corrections for inclusion in the FITS headers and publishes
metrics such as instantaneous seeing and sky transparency, which the scheduler
uses to determine the sequence of upcoming observations (\citealp{farina2018}).

\begin{figure*}
  \centering
  \includegraphics[width=\textwidth]{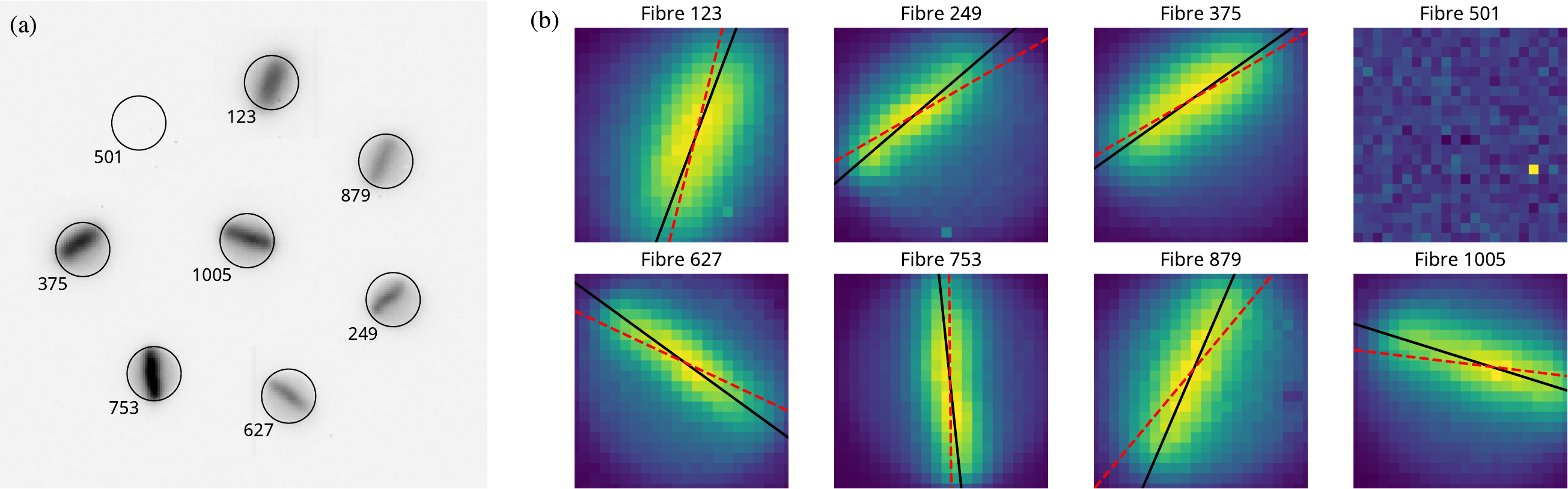}
  \caption{(a) Calibration image taken on 2022 November 9, with the telescope
  slowly offsetting in +Dec and the stars moving across the AG FOV, thus
  producing streaks on the detector. The image presented here is not the
  original (since the fibres would be too small to see), but the remapped image
  used for real-time display, see \figref{fig:mosccd}.
  (b) The streaks produced in each fibre are analysed by fitting a line through
  them, from whose slope the fibre orientation can be determined after
  accounting for the sky PA and for the detector orientation and flip. For each
  fibre we plot the old slope, stored in the AG software (dashed red line) and
  the new slope, calculated from the fit (solid black line).}
  \label{fig:streaks}
\end{figure*}

As mentioned in \secref{mosag:autoacq}, the full astrometric calculations from
mean ICRS coordinates to plate coordinates (\appref{app:coords}) are performed
for every guide frame, taking into account the current time, differential
refraction, and any potential desired offsets in the case of dithered
observations. Similar to the LIFU acquisition (\secref{lifuag:autoacq}) and
guiding (\secref{lifuag:guiding}), these calculations ultimately update the
expected positions of the guide stars such that the optical axis of the
telescope keeps pointing at the field coordinates \textsc{fldra} and
\textsc{flddec}. These are calculated by the positioner when placing the science
fibres on the plate, such that pointing at the field coordinates ensures that
the science targets fall inside their respective fibres.

In the case of dithered MOS/mIFU observations, the OCS simply offsets
\textsc{fldra} and \textsc{flddec} according to the desired pattern; the AG has
no concept of dithering: by simply recalculating the expected positions of the
guide stars according to the current \textsc{fldra} and \textsc{flddec}, we
ensure that the AG guides on the dithered position, and that all science targets
are shifted by the appropriate amount inside the science fibres.

In the case of WTL OBs (see \secref{intro:weavedesign}), the AG system must be
able to resume closed-loop guiding immediately after blind telescope offsets of
up to $\sim 15$\textdegree, using a different set of guide stars at each field
position. The ability to rapidly reacquire the guide stars and re-establish
stable guiding with minimal overhead is therefore a key capability of the AG
system and underpins the practical feasibility of such observing strategies.

\subsection{Guide fibre calibrations}\label{mosag:calibs}
\subsubsection{Determining the guide fibre positions on the CCD}\label{calibs:fibrepos}
Since the MOS AG CCD must be able to receive light from either of the two sets
of guide fibres, physical contact between the guide fibre ends and the detector
is not an option. Instead, a mechanism using a stepper-motor-driven linear slide
positions the correct set of guide fibres in front of the detector whenever the
positioner tumbles, and the images of the guide fibres are transferred optically
to the detector through a system of lenses, as shown in inset (c) of
\figref{fig:agdiagram}.

The exact positions of the fibres on the detector depend therefore on the plate
in use and on the exact position of the exchange mechanism's stepper motor. In
practice, any kind of work related to the slide exchange or to the detector can
shift the position of the fibres on the CCD. During commissioning, small shifts
of the order of a few pixels, likely related to flexure, have even been observed
within the course of an observing night, as the telescope slewed to different
elevations. It is therefore important to have a fast, accurate, and automatic
way of determining the fibre positions on the CCD, since the acquisition and
guiding algorithm (\secref{mosag:autoacq}) depends critically on knowing where
the fibres are on the detector.

Whenever the fibre calibration routine is triggered, the AG turns on a helium
calibration lamp in the WHT calibration unit, and takes an exposure long enough
to uniformly illuminate the fibres without saturation. It then analyses the
profiles of the fully illuminated fibres in order to determine the centroid of
each fibre, and presents the user with a comparison between the old values
(stored in the AG configuration files) and the newly measured values; the user
can then decide whether to ingest the new values or to ignore them. The entire
calibration procedure takes $\sim 30$~s, and is automatically performed at the
beginning of every MOS/mIFU OB, just before acquisition starts.

\subsubsection{Determining the guide fibre orientations}\label{calibs:fibreorient}
One end of the guide fibres is placed on the plate by the positioner when
configuring a field, while the other end feeds into the guide slide, where the
image is relayed on to the AG detector. At the plate end, the rotation of the
guide fibres is physically constrained: it is zero when they are placed
radially, and can reach several degrees when placed non-radially; the exact
value is calculated and published by the positioner during the plate
configuration. At the detector end, however, the rotation is not physically
constrained, so it was essential to measure it on-sky; this orientation has also
on occasion changed during detector alignment or guide-slide adjustment work.

Accurate measurements of the fibre orientations at the detector end are crucial,
because the pixel displacements of the guide stars from the fibre centres, used
for computing corrections during both acquisition and guiding, can only be
converted to displacements in plate coordinates after accounting for the fibre
orientations.

The simplest way of measuring the fibre orientations is to take an exposure
where stars are visible in all the fibres, then offset the telescope by a small
amount (e.g., 1~arcsec) in a given direction (RA or Dec), and then take another
exposure. The direction in which each star moves on the CCD, after accounting
for the sky PA, the detector orientation, and any possible flip, then yields the
orientation of each fibre. In practice, the various sources of noise in the
image, the limited FOV of the guide fibres, and seeing effects make such single
readings unreliable, as even an error of one pixel orthogonal to the movement
can amount to over 10\textdegree\ in orientation.

An improvement to this method is to adopt a statistical approach and to repeat
the individual measurements, for example by offsetting the telescope repeatedly,
which should average out the random components in the measurements and reveal
any real aberration.

A faster method that we devised, however, is to acquire the stars in the fibres
and then offset the telescope so that the stars go out of the FOV; we then start
exposing and at the same time offsetting the telescope to slowly move the stars
across the FOV, while the camera keeps integrating. This results in streaks
across the FOV that can be fitted with a line (see \figref{fig:streaks}). By
performing various such measurements, we confirmed that the standard deviation
given by this method is less than 1\textdegree, which is sufficient for our
purpose. Apart from its excellent accuracy, this method also has the advantage
of being extremely fast, as only one measurement/exposure is necessary (as
opposed to the statistical approach mentioned before). The entire calibration is
automated by a script that takes exposures and slews the telescope, fits lines
through the streaks in all the fibres, corrects for sky PA and camera
orientation and flip, and updates the configuration file with the new values.
Since the streaks cross the full extent of the fibres, the direction of motion
cannot be deduced based on the resulting image alone, as there are always two
solutions that differ by 180\textdegree\ for the fitted slope: the algorithm
always selects the solution that is closest to the old one (which means that the
very first values were determined by hand, by picking one of the solutions,
offsetting the telescope, and checking whether the direction of motion was
accurately reflected in the de-rotated fibre images).

Once guide stars are acquired inside the fibres, the automated calibration 
procedure takes $\sim 2$ minutes, and therefore can be performed as often as
necessary, with very little overhead.

\subsection{Focusing the telescope using multiple guide fibres}\label{mosag:multifocus}
In MOS/mIFU modes, acquisition and guiding are performed using coherent fibre
imaging bundles that are deployed in the telescope focal plane alongside the MOS
science fibres (or the mini-IFUs, respectively). At the telescope focus, each
guide fibre forms an image of the sky on the front surface of a coherent bundle.
This image is transmitted to the instrument enclosure, where it appears on the
output face of the bundle and is re-imaged onto the AG detector by a fast
optical relay comprising a pair of f/1.4 lenses, see inset (c) of
\figref{fig:agdiagram}.

The guide fibres at the telescope focal plane are designed to be parfocal with
the MOS science fibres (or the mini-IFUs). Consequently, achieving optimal focus
for the guide stars at the telescope focal plane ensures that the science
targets are also optimally focused.
The optical design defines a unique best-focus position, but the observed image
size is dominated by atmospheric seeing. As a result, accurate focusing of the
telescope itself is a prerequisite for all subsequent acquisition and guiding
operations, as well as for the characterization of internal focus mechanisms.

\subsubsection{Measuring the telescope focus}
\begin{figure}
  \centering
  \includegraphics[width=\columnwidth]{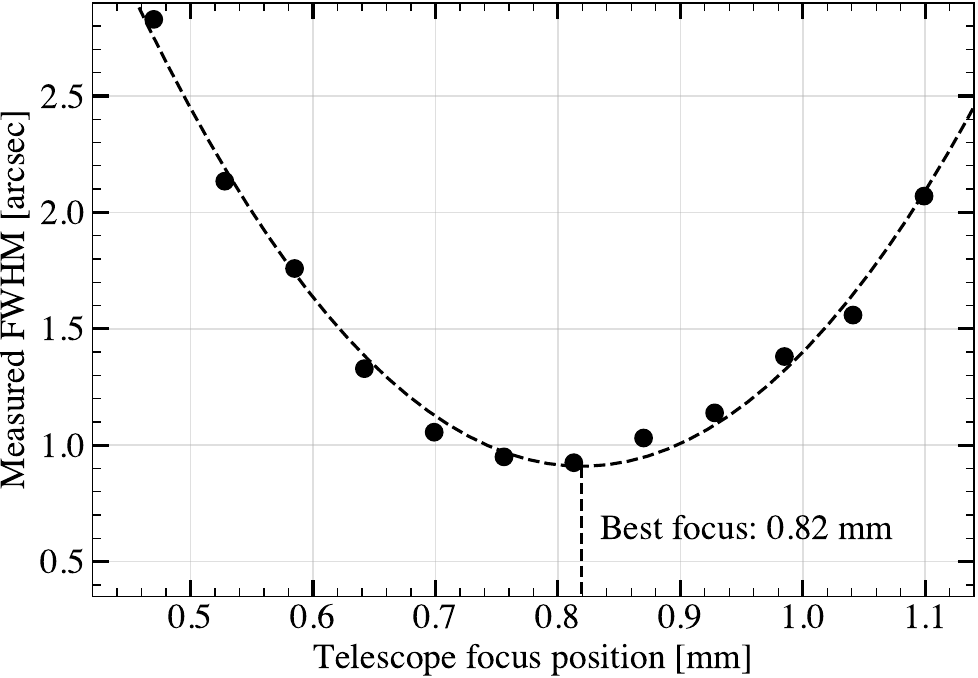}
  \caption{Example of a typical focusing curve using data taken on 2025 August
  10. The measured FWHM as a function of the telescope focus position (filled
  black circles) tends to be well represented by a parabola (dashed curve),
  whose vertex is a good estimation of the `best focus'.}
  \label{fig:focusPlot}
\end{figure}

Provided that the AG optical relay is already set to a reasonably good focus
(which can later on be refined, see \secref{sec:internalfocus} below),
both theoretical considerations and empirical tests show that, in the
seeing-dominated regime, the telescope focus produces a parabolic variation in
the measured image size on either side of the optimum focus. The vertex of this
parabola provides a robust estimate of the best telescope focus
(\figref{fig:focusPlot}). Although the optical design introduces mild asymmetry
in the curve, on-sky data confirm that the vertex is a stable indicator except
under strongly variable seeing. Such conditions can produce large residuals near
the centre of the fit or a discrepancy between the vertex and the position of
minimum image full width at half maximum (FWHM), prompting the observer
to repeat the measurement.

A practical limitation arises when the stellar images become strongly defocused
during the focus sweep. As the spot size approaches the physical diameter of the
guide fibre, the recorded PSF becomes dominated by the fibre aperture rather
than the true optical blur. In this regime the measured FWHM ceases to vary
smoothly with focus, and the parabolic fit becomes less reliable, especially for
guide stars that are not well-centred inside their respective fibres. In
addition, the focus sequence itself typically spans $\sim 4$ min, during which
small but non-negligible telescope drifts can occur. Any such drift changes the
sampling of the stellar profile across the fibre face, introducing spurious
variations in the measured FWHM that broaden the scatter around the fitted
curve. These effects are usually modest but can degrade the accuracy of the
inferred best-focus position, especially under poor seeing or when guide stars
lie close to the fibre edge. Proper interpretation of the autofocus results
therefore requires awareness of these limitations and, where possible, repeating
the focus sequence when residuals indicate that either effect may have been
significant.

For each focus run, summary statistics, diagnostic plots, and interactive web
pages are generated for all individual guide stars. The numerical results are
archived to enable long-term monitoring.

\subsubsection{Measuring the focal plane tilt}
In MOS mode, the guide stars used for autoguiding are typically well distributed
across the field. We exploit this to determine not only the global telescope
focus but also its variation across the focal plane. We developed a dedicated
wrapper script that first calls the standard
single-star autofocusing routine to obtain a sequence of images taken at
different focus positions. For one guide fibre, the script fits a Gaussian
profile to the stellar image in each frame, constructs the FWHM--focus curve,
and determines the best focus from a parabolic fit. The same set of images is
then reused for all remaining fibres, whose analysis runs in parallel and adds
only a few seconds to the overall procedure.

\begin{figure}
  \centering
  \includegraphics[width=\columnwidth]{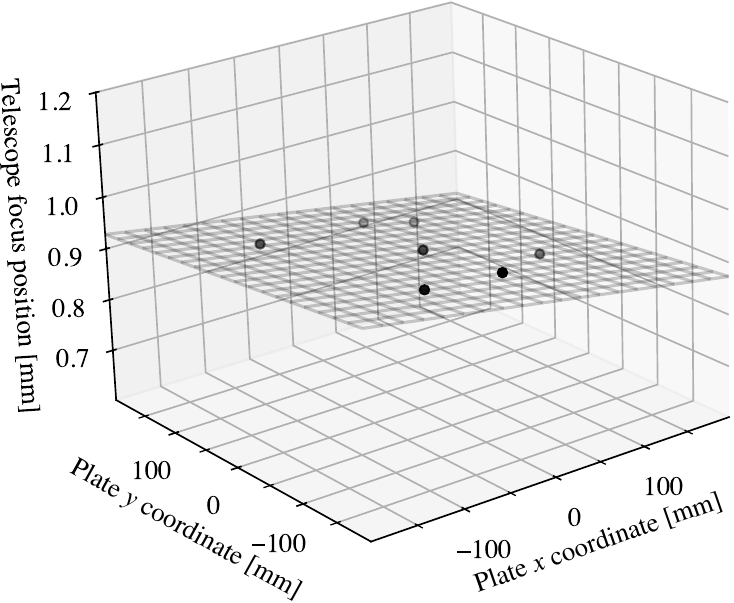}
  \caption{Measured best-telescope focus positions for individual guide fibres
  (black points) shown as a function of their physical locations on the MOS
  plate, using data taken on 2025 November 14. A plane is fitted to these values
  (wireframe), revealing a tilt of the focal surface with respect to the optical
  axis of $0.019$\textdegree, corresponding to an offset of 65~$\umu$m at the
  edge of the plate.}
  \label{fig:fittedPlane}
\end{figure}

The best-focus positions derived for all guide fibres are combined with their
known physical coordinates on the MOS plate to fit a plane
(\figref{fig:fittedPlane}). This provides an estimate of the tilt of the focal
plane relative to the optical axis. The measured tilts typically lie between 0
and 0.03\textdegree, and are repeatable even when the residuals of the plane fit
are small ($\sim 10$ $\umu$m), indicating that the observed tilt reflects
genuine mechanical alignment rather than noise. These values are also consistent
with the detailed metrology conducted at the WHT prior to the final top-end
assembly (\citealp{hughes2022b}).

The quality of the plane fits depends primarily on atmospheric seeing and on how
well the guide stars are centred within their fibres. For runs using at least
four guide fibres, the rms scatter about the fitted plane is below 10~$\umu$m in
15 per cent of cases, below 20~$\umu$m in 53 per cent, and below 30~$\umu$m in
79 per cent. Poorer fits usually arise when the guide star lies close to a fibre
edge (where accurate profile fitting becomes difficult) or when the telescope
focus sweep does not bracket the true focus, producing a biased parabolic fit.

\subsubsection{Measuring the internal AG focus}\label{sec:internalfocus}
Once the telescope focus has been established, the same parabolic focus-curve
method can be applied to characterize the backfocus of the internal MOS AG
focusing mechanism. 

This design of the AG optical relay minimizes light-transfer losses and has
demonstrated exceptional sensitivity, enabling acquisition and guiding on stars
as faint as 19.4~mag, well beyond the original specifications. The large
relative aperture, however, results in an extremely shallow depth of focus: the
lens back-focal distance must be set with a precision of $\sim$25~$\umu$m. Such
accuracy cannot be judged reliably from fibre-end images alone, as these lack
well-defined reference marks or crosshairs. To allow repeatable and quantitative
focusing, the front lens of the AG relay is mounted on a high-precision actuator
under PLC control, as shown in inset (c) of \figref{fig:agdiagram}. The
focus-measurement procedure described above is used to characterize both the
telescope focus and the internal MOS AG focusing mechanism, differing only in
which optical element is adjusted. In the case of the latter, the internal focus
mechanism is driven through a series of discrete positions while the telescope
focus is held fixed.

Despite the precision of the focusing procedure, the minimum FWHM measured on
the MOS AG camera rarely falls below $\sim$0.9 arcsec, often exceeding
nearly-simultaneous differential image motion monitor (DIMM; e.g.,
\citealp{sarazin1990}) or FPI seeing measurements by $\sim$20--25 per cent. The
autoguider optics themselves cannot account for this discrepancy: independent
measurements show they contribute an effective broadening of only $\sim$0.4
arcsec, which, when combined with typical seeing, explains at most $\sim$15 per
cent of the increase. The remaining difference is likely to originate from other
misalignments, including small variations in the heights of the MOS guide fibres
within their alignment block at the detector end. A mechanical solution to
improve the alignment of the system is currently under investigation.

\medskip

\noindent The same focusing procedure described in this section, including the
construction and fitting of the parabolic focus curve, is used consistently for
MOS/mIFU, LIFU, and FPI-based focusing, differing only in the source of the
image data used to compute the focus.


\section{The simulation mode}\label{sec:sim}
A fully fledged simulation mode for the autoguider proved essential for its 
successful and efficient development. Because the AG interacts with several
complex subsystems, most notably the TCS and the guide cameras, these components
also needed to be simulated at an adequate level of fidelity. Building such an
end-to-end simulation framework required a substantial development effort, but
it was indispensable: in the case of WEAVE, the autoguider could not be fully
tested prior to first light, and any major issues discovered at that stage would
have risked delaying the entire project. 

By implementing the simulated system described in this section, we were able to
fully test the WEAVE AG well in advance of the instrument completion, and to
gain confidence in its fitness for purpose, stepping into commissioning with the
assurance that there would be no show-stoppers on the part of the AG. This was
indeed the case: the AG guided successfully during its first light (2022 June
16), and has ever since, without any major software problems, which fully
justifies the effort invested in the simulation mode. Moreover, even today, as
WEAVE approaches routine operations, we still use the simulation mode on a
regular basis for reproducing and fixing faults, and testing improvements and
new features, with the confidence that once they work off-sky in simulation
mode, the changes will work as expected on-sky.

\subsection{The simulated telescope control system}\label{sim:tcs}
In preparation for the arrival of WEAVE, the WHT TCS hardware
(\citealp{boksenberg1985,jones1986}) was upgraded with PLCs in 2018, and is now
controlled by a Linux-based `TCS bridge computer' that also acts as the
interface between the TCS and the rest of the subsystems (see
\figref{fig:ocssoftware}); however, the OpenVMS-based machine that receives the
serial guide correction packets from the autoguider (\citealp{laing1993b}) and
performs the astrometry calculations (\citealp{laing1993a}) for pointing and
tracking (the `VMS TCS') has so far remained largely unchanged.

The VMS TCS already provides a simulation mode, which runs the actual Fortran
code on a DEC Alpha machine without controlling the telescope hardware. However,
it is relatively time-consuming to set up, and requires a dedicated, spare Alpha
machine, as well as a spare TCS bridge computer and spare observing control
system machine to interface with it. It also lacks several desirable features
for simulating acquisition and guiding, as described below. We therefore chose
to reimplement a simulated WHT TCS in Python, which fulfils the roles of both
the bridge computer and of the VMS TCS.

At its core, a fourth-order Runge--Kutta integrator provides simulated tracking
by time-integrating the same differential equations for zenith distance, azimuth
and parallactic angle as the real VMS TCS (\citealp{laing1993a}). On top of it,
the same interface provided by the TCS bridge computer allows the user to slew,
change the rotator angle, start and stop guiding, etc., as well as collect
TCS-related FITS header keywords, while an emulated serial device to which the
simulated TCS listens allows it to receive (and apply) guide correction packets. 
For all intents and purposes, for any external software that communicates to the
TCS during normal operations, the simulated TCS is identical to the real one.

The starting point for the time integration can be set to an arbitrary value,
which means that simulations can be run for past and future dates -- use cases
include simulating an observation for the current night during daytime,
repeating an observation from a previous night while addressing a fault, and
repeatedly acquiring the same field under the exact same conditions. This
feature is not available in the VMS TCS simulation mode, but has proven
invaluable for debugging.

In addition, the conditions of specific on-sky observations (including pointing
coordinates, rotator angle, fibre setup, and weather conditions) can be
reproduced from a single line logged by the AG to the system log. We found this
feature essential for handling guiding-related faults that occurred during
operations (such as problems with the LIFU finding chart, LIFU acquisition in
very sparse or very crowded fields, etc.) and reproducing them in a controlled
environment. Additionally, since most of the logged items are propagated from
the positioner to the AG, this feature has also helped detect issues in the
positioner software unrelated to the AG.

\subsection{The simulated WEAVE AG camera}\label{sim:agcamera}
The purpose of the simulated AG camera is to generate images that closely
reproduce what the real camera would record on-sky at the telescope coordinates 
provided by the simulated TCS, under a configurable set of observing conditions.
Although UltraDAS includes a simulation mode, it is intended primarily for
testing UltraDAS itself: the simulated images are static frames read from disk,
they do not respond to telescope motion, and running UltraDAS in simulation mode
requires the full infrastructure of the WHT observing system. A variety of
high-fidelity image simulators, both general-purpose and telescope-specific,
exist in the literature, such as SkyMaker (\citealp{bertin2009}), UF\textsc{ig}
(\citealp{berge2013}), GalSim (\citealp{rowe2015}), PhoSim
(\citealp{peterson2015,burke2019}), STIPS (\citealp{gomez2024}), SphereX
(\citealp{crill2025}), Open\-Universe2024 (\citealp{alarcon2025}), etc. While
they focus primarily on PSF realism, galaxy morphology, and detailed detector
physics, none of these tools provide the instrument-specific combination of
high-precision sky-to-detector astrometry, fibre-based guider imaging, and
closed-loop  interaction with a simulated telescope control system required for
testing the WEAVE AG algorithms. Our simulator therefore targets a different
niche: reproducing, in real time, both the astrometric behaviour and the
feedback dynamics essential for validating acquisition, guiding, and TCS
interaction.

In MOS mode, the astrometric parameters of the (up to eight) guide stars are
supplied by the positioner. In LIFU mode, the simulator retrieves the stars
within the AG FOV from the Gaia catalogue. Extended sources (such as galaxies)
and moving objects (such as asteroids and comets) are not included in the
simulation at this stage. In both cases, the simulated camera reads the current
telescope pointing from the simulated TCS (thereby incorporating dithering
offsets and pointing errors) and performs the full astrometric transformation
from mean ICRS coordinates to detector coordinates (\appref{app:coords}). This
includes optical distortion and differential refraction, enabling
point-for-point accurate placement of stars on the detector. The flux
distribution of each star is then generated using its catalogue magnitude, a
configurable seeing value (or one randomly drawn from a configurable range), and
a selectable configurable PSF profile (e.g., Gaussian or Moffat).

\begin{figure}
  \centering
  \includegraphics[width=\columnwidth]{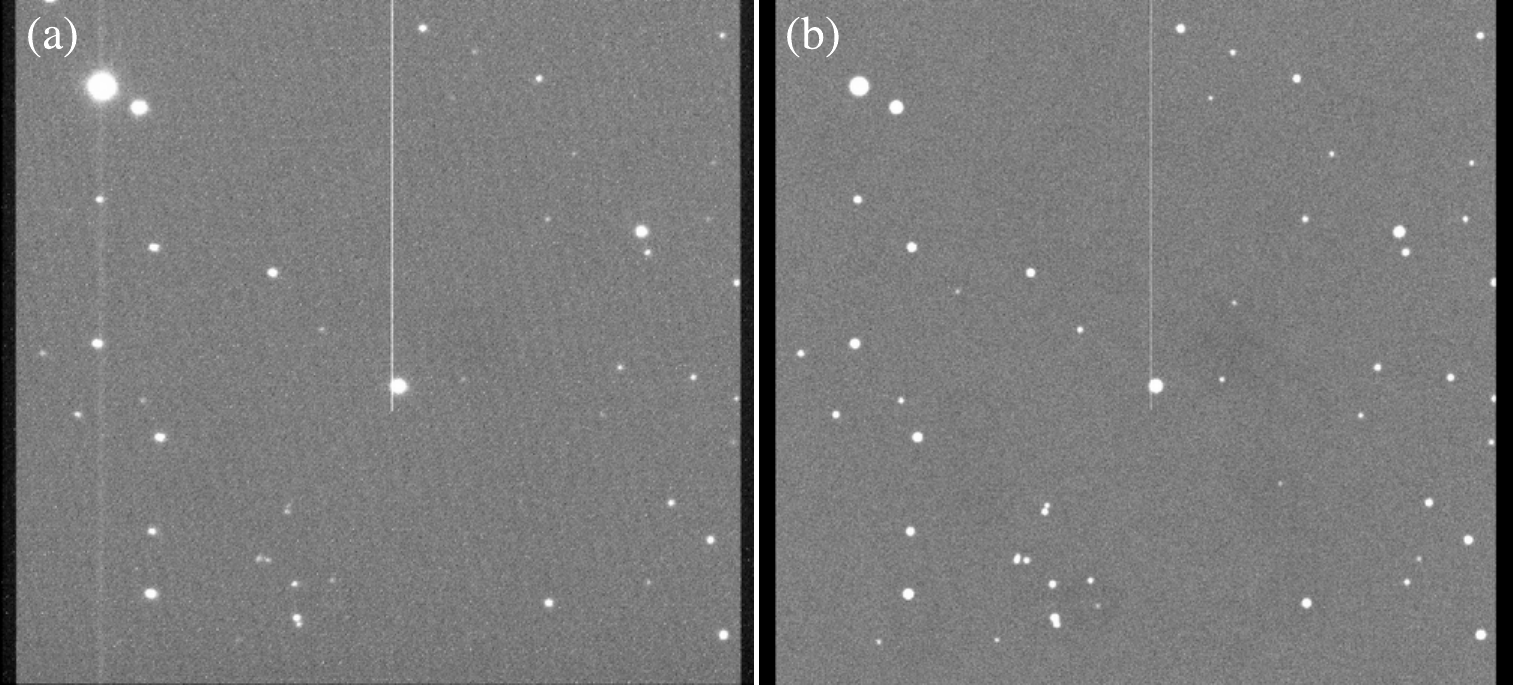}
  \caption{(a) A LIFU AG acquisition image of WEAVE OB 6073 taken on 2025
  September 13.
  (b) A synthetic image of the same acquisition field, generated with the
  simulated WEAVE AG camera using the Gaia catalogue, and based solely on a
  single line of the logged output from the night-time observations.
  Equally realistic images can be generated for any celestial coordinates, and
  for both the MOS and the LIFU AG.}
  \label{fig:synth}
\end{figure}

Instrument- and mode-specific behaviour is also reproduced. In MOS mode, the
simulator generates fibre images of the correct size and at the correct
positions, and optionally adds a configurable sky background to each fibre,
enabling controlled tests of AG performance under bright-sky or thin-cloud
conditions. In LIFU mode, detector artefacts such as bright columns, bad pixels,
and flat-field structures are included, allowing the AG's real-time reduction
pipeline to be tested under realistic detector pathologies (see
\figref{fig:synth}).

As with the simulated TCS, the camera simulator exposes an interface identical
to that of the real UltraDAS cameras. Together, these features make the
simulated AG camera an end-to-end, realistic test environment that can fully
replace the real system for development, debugging, and automated testing
without requiring telescope time or access to the observatory infrastructure.

\begin{table*}
\centering
\caption{Summary of WEAVE acquisition and guiding performance for LIFU and
MOS/mIFU observing modes, using on-sky data from 2024--2025 and selected as
described in \secref{results:methods}.}
\label{tab:ag_summary}
\begin{tabular}{lcc}
\hline
Metric & LIFU & MOS / mIFU \\
\hline

\multicolumn{3}{l}{\textit{Acquisition performance}} \\
Acquisition method & Pattern recognition (asterism hashing) & Multi-fibre centroiding \\
Acquisition geometry & Off-axis imager & Distributed coherent guide fibre bundles \\
Typical cadence & 25--30 s & 20--25 s \\
Total duration, median (90\textsuperscript{th} percentile) & 53 (128) s & 76 (309) s without FPI; 309 (756) s with FPI \\
Number of iterations, median (90\textsuperscript{th} percentile) & 2 (6) & 7 (13) \\
Translational error, median (90\textsuperscript{th} percentile) & 0.19 (0.44) arcsec & 0.16 (0.31) arcsec \\
Rotation error, median (90\textsuperscript{th} percentile) & N/A & 0.53 (1.55) millideg \\
Rotation correction during acquisition & Yes & Yes \\
Acquisition success rate & 96\% & 72\% (without FPI); 100\% (with FPI) \\
Dominant acquisition failure mode & Sparse field / cloud & No stars in fibres \\
Fallback strategy & Sky PA change (sparse field) / repeat later (cloud) & FPI acquisition or spiral search \\

\hline
\multicolumn{3}{l}{\textit{Guiding performance}} \\
Typical guiding cadence & 8--10 s & 8--10 s \\
Guiding rms, translational, median (90\textsuperscript{th} percentile) & 0.24 (0.36) arcsec & 0.21 (0.28) arcsec \\
Guiding rms, rotation, median (90\textsuperscript{th} percentile) & N/A & 0.77 (1.71) millideg \\
Guiding requirement & 0.3-arcsec rms & 0.3-arcsec rms \\
Requirement met & 75\% of OBs & 96\% of OBs \\
Rotation correction during guiding & No (under investigation) & Yes \\

\hline
\multicolumn{3}{l}{\textit{Operational robustness}} \\
Number of OBs (exposures) & 592 (1872) & 149 (447) \\
DIMM seeing, median (range) & 0.81 (0.35--4.56) arcsec & 0.76 (0.36--2.28) arcsec \\
Integrated airmass, median (range) & 1.13 (1.00--2.09) & 1.14 (1.02--1.98) \\
Guide star magnitudes, median (range) & 17.08 (14.51--20.04) & 15.75 (14.53--19.39) \\
Guiding failure rate & 0\% & 0\% \\
Differential refraction handling & Per science exposure (before Dec 2025) & Per AG frame \\

\hline
\end{tabular}
\end{table*}

To our knowledge, no existing guider-camera simulator provides realistic,
on-the-fly images while maintaining full compatibility with the instrument
control system. Given the potential wider applicability of such a tool (e.g.,
for testing guiding algorithms, validating reduction pipelines, developing
commissioning procedures, or enabling remote software development without access
to telescope infrastructure), we have created a fully functional,
instrument-agnostic version of the camera simulator. This modified version,
decoupled from WEAVE-specific components, is hereby released under the MIT
license as \textsc{siren} (Sky Image Rendering ENgine) on
GitHub\footnote{\href{https://github.com/egafton/siren}{https://github.com/egafton/siren}}.

\subsection{The simulated autoguider}\label{sim:ag}
Because both the simulated TCS and the simulated AG camera expose the same
interfaces as their real counterparts, the AG software itself requires no
dedicated simulation mode (see also the software processing layer in
\figref{fig:agdiagram}, which shows how the simulated camera and TCS are
essentially `plug-in' replacements for
UltraDAS and the real TCS, respectively, from the point of view of the AG).
The AG code runs unmodified, whether on-sky or
entirely within the simulated environment. This capability allowed us to
identify and resolve issues well in advance of commissioning, and to verify its
compliance with all interface, GUI, computational, and guiding-correction
requirements long before telescope integration. As a result, during first light
the AG was running exactly the same code that had been exercised and validated
for months. Since then, the simulation infrastructure has continued to enable us
to reproduce and resolve night-time issues in a controlled daytime environment.

Beyond its role in software verification and algorithm development, the
simulation mode proved to be a critical operational and organisational asset.
The availability of a fully functional simulated WEAVE environment (including
acquisition and guiding, OCS interactions, and graphical user interfaces)
enabled the astronomers and telescope operators to use it as a training platform
for more than a year prior to on-sky commissioning. This allowed users to
rehearse complete observing workflows, become familiar with the AG user
interfaces, and provide structured feedback on functionality and usability in a
low-pressure environment, without the constraints and risks associated with
night-time commissioning. As a result, several interface refinements and
additional features were implemented before first light, and communication
between the engineering and science teams was substantially improved through
shared use of a common, realistic system.


\section{Results}\label{sec:results}

\begin{figure*}
  \centering
  \includegraphics[width=\textwidth]{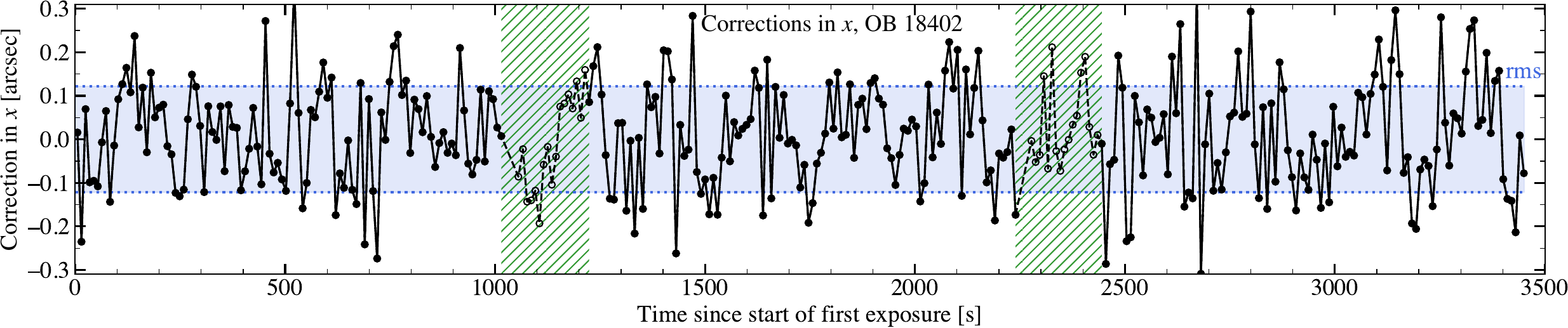}
  \caption{Autoguider corrections in $x$ over the course of a typical WEAVE
  science OB with 3 exposures of 1020~s each, observed on 2025 July 7. The
  frame-by-frame AG corrections applied during the science integrations (filled
  black circles, joined by solid black lines) are stored in the FITS headers,
  but the corrections applied while the cameras are reading out (open black
  circles, joined by dashed black lines) are not -- for this plot, they have
  been retrieved from the AG system logs and are additionally indicated by the
  green hatched regions. The shaded blue area indicates the expected amplitude
  of typical guiding fluctuations, as a symmetric band at $\pm{\rm rms}_x$
  (dotted blue lines), as given by \eqreft{eq:rms}. For this OB,
  ${\rm rms}_x=0.12$~arcsec.}
  \label{fig:oneline}
\end{figure*}

This section presents the measured performance of the WEAVE AG system based on
the extensive dataset accumulated during commissioning and the first year of
routine survey operations. The analysis uses only internal WEAVE telemetry
(specifically, the AG binary tables embedded in all science and calibration FITS
files, autoguider system logs, autofocusing summaries, and the WHT observing
log) and therefore reflects the behaviour of the system under genuine on-sky
conditions and across the full range of observing programmes.

A quantitative overview of acquisition and guiding performance for both LIFU and
MOS observing modes, summarizing the results presented in this section, is
provided in \tabref{tab:ag_summary}.

\subsection{Dataset and methods}\label{results:methods}
All WEAVE science files contain per-frame autoguider telemetry, including the 
timestamp of each AG frame and the measured guiding offsets in $x$ and $y$, and
(in the case of MOS) the applied correction in rotation. For this work we
analysed 2319 exposures taken with the WEAVE RED camera between 2024 January 1
and 2025 November 18, corresponding to $\approx 657$ hours of live guiding time
and encompassing over 290\,000 AG corrections applied to the TCS. Of these
exposures, 1872 were taken in LIFU mode, 240 in MOS A, and 207 in MOS B. Within
the given time interval, we selected all of the exposures with an integration
time of 1020~s (the standard for WEAVE science in both MOS and LIFU) that were
part of a science OB (as opposed to engineering, calibration, or test exposures)
with at least 3 exposures and that had guiding data in the FITS headers. No OBs
from the given time interval that fulfilled these criteria have been excluded.

The science exposures we analysed were part of a total of 741 WEAVE science OBs,
of which 592 (80 per cent, totalling 1872 exposures) were acquired with the LIFU
AG, and 149 (20 per cent, totalling 447 exposures) with the MOS AG (in either
MOS A or MOS B modes; no mIFU science OBs have been observed so far).

Because the BLUE and RED cameras acquire science exposures simultaneously and
always with identical exposure times, their AG data are intrinsically equivalent
for the purpose of analysing guiding performance. For this reason, and without
loss of generality, all guiding-performance results in this section are based
exclusively on RED exposures.

The exposures analysed here span the following ranges of parameters: integrated
airmasses from $1.00$ to $2.09$ (or, conversely, zenith distances from
$1.07$\textdegree\ to $63.25$\textdegree), median DIMM seeing from $0.35$ to
$4.56$~arcsec, and guide star magnitudes from $14.5$ to $20.0$~mag.

For each OB we extracted:

\begin{enumerate}
\item the frame-by-frame guiding residuals in $x$, $y$ and (in the case of MOS)
rotation, defined as the difference between the measured and the expected
positions of the guide stars (filled black circles in \figref{fig:oneline});
\item the root-mean-square (rms) values over the entire OB, computed as
\begin{equation}\label{eq:rms}
  {\rm rms}_x=\sqrt{\frac{1}{N}\sum_{i=1}^{N} {x_i}^2},\quad
  {\rm rms}_y=\sqrt{\frac{1}{N}\sum_{i=1}^{N} {y_i}^2},\quad
  {\rm rms}_\theta=\sqrt{\frac{1}{N}\sum_{i=1}^{N} {\theta_i}^2},
\end{equation}
\noindent where $N$ is the number of autoguider samples during the given OB, and
$x_i$, $y_i$, $\theta_i$ are the corrections in $x$, $y$, and (in the case of
MOS) rotation, respectively, of the $i^{\rm th}$ sample (blue line in
\figref{fig:oneline}). Where the total translational rms over the duration of
an OB was necessary, we computed it from the previous equations as
\begin{equation}\label{eq:rmsxy}
{\rm rms}_{xy} = \sqrt{{{\rm rms}_x}^2 + {{\rm rms}_y}^2};
\end{equation}
\item the cumulative offsets over the entire OB, computed as
\begin{equation}\label{eq:cumulative}
X_n = \sum_{i=1}^n x_i,\quad
Y_n = \sum_{i=1}^n y_i,\quad
\Theta_n = \sum_{i=1}^n \theta_i,
\end{equation}
\noindent where $x_i$, $y_i$, $\theta_i$ are the measured offsets at guiding
step $i$, and $n$ taking values from $1$ to $N$;
\item the real discrete Fourier transform (see, e.g., \citealp{press1992})
$H_x$, etc. of the AG corrections in $x$, etc. linearly-interpolated on to a
uniform time grid with 10-s spacing, $\bar{x}$,
\begin{equation}\label{eq:rfft}
H_x(f_n) = \sum_{k=0}^{N-1} \bar{x}_k e^{-2\pi i k n / N},\quad n=0,\ldots,N-1,
\end{equation}
\noindent and the corresponding power spectral density,
\begin{equation}\label{eq:psd}
{\rm PSD}_x(f_n) = \left|H_x(f_n)\right|^2.
\end{equation}

\end{enumerate}

\begin{figure*}
  \centering
  \includegraphics[width=\textwidth]{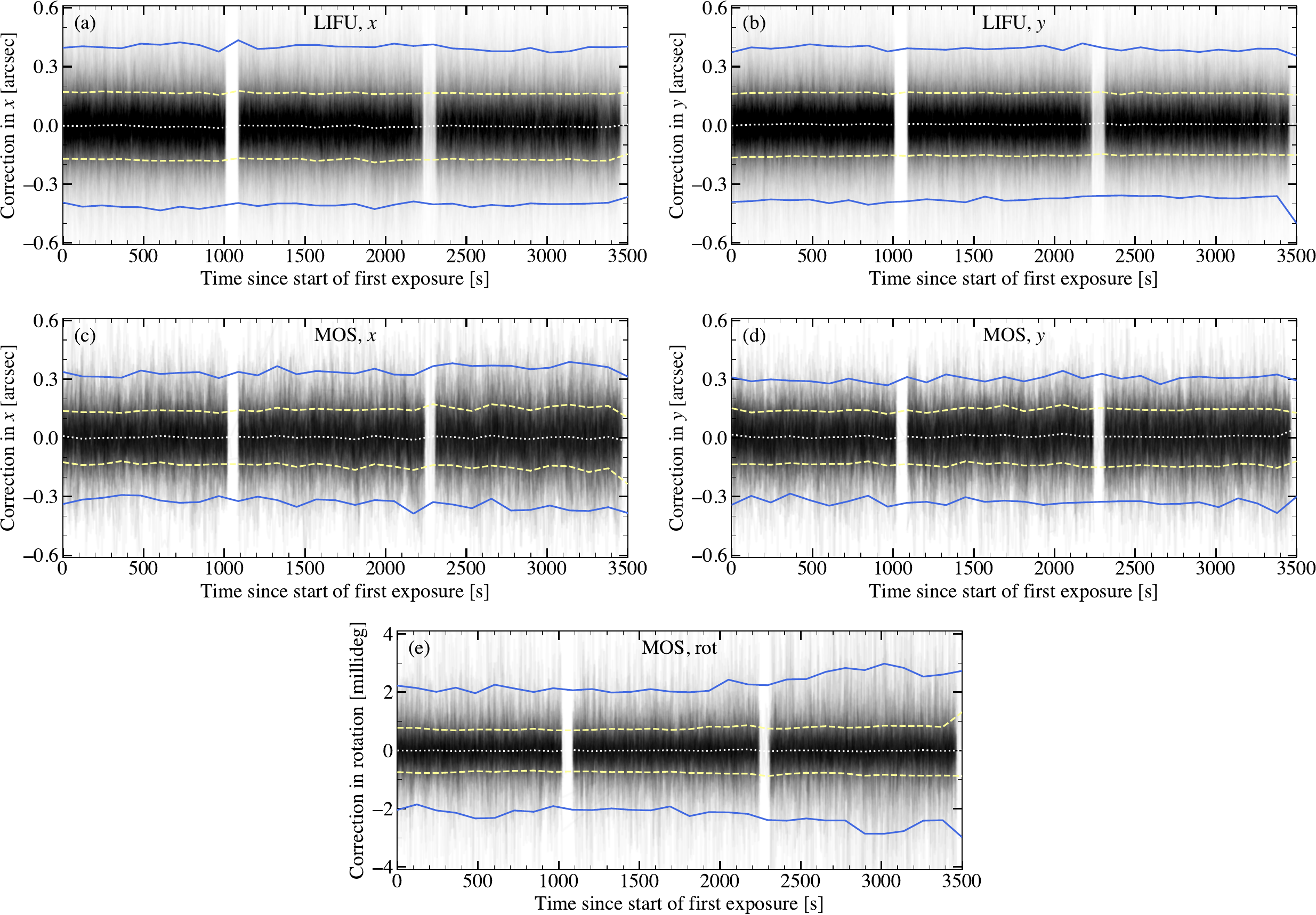}
  \caption{Alpha-blended ensemble of all AG corrections.
  (a) and (b): LIFU AG corrections in $x$ and $y$, respectively, applied during
  1872 exposures from 568 LIFU OBs.
  (c), (d) and (e): MOS AG corrections in $x$, $y$ and rotation, respectively,
  applied during 447 exposures from 149 MOS OBs. Each thin grey line corresponds
  to the science exposures within a single OB, plotted with the time elapsed
  since the start of the first exposure on the $x$-axis. The dotted white lines
  show the median corrections, while the dashed yellow ($1\sigma$) and solid
  blue ($2\sigma$) lines enclose  $\sim 68$ and $\sim 95$ per cent of all
  corrections, respectively, and all are calculated for bins of 120-s width.
  Faint trajectories indicate rare outlier events that may be caused by tracking
  and/or rotator instabilities, or by external effects such as windshake.}
  \label{fig:guidefull}
\end{figure*}

The guiding residuals analysed in this section reflect the behaviour of the full
end-to-end acquisition and guiding system rather than the performance of the AG
software in isolation. In both MOS and LIFU modes, the measured guiding accuracy
is influenced not only by centroiding precision and AG control-loop behaviour,
but also by the execution fidelity of the TCS, including its internal
closed-loop response, backlash compensation and pointing repeatability, and by
effects such as hysteresis and mechanical flexure. In MOS mode, additional
contributions arise from the accuracy and repeatability of the fibre positioner,
and from residual fibre placement errors. As a result, the residual and rms
values reported below should be interpreted as system-level performance metrics
that quantify the stability and robustness of the combined POS--AG--TCS
mechanical system under real operating conditions, rather than as direct
measures of individual subsystem performance.

\subsection{Acquisition statistics}\label{results:acquisition}
Of the 149 MOS OBs, 108 (72 per cent) were acquired solely by the MOS AG, while
for the other 41 (28 per cent) no stars were initially visible in the fibres, so
the operator had to deploy the FPI and acquire two stars manually
(\secref{mosag:fpiacq}) before proceeding with the automatic MOS AG acquisition.
The spiral search (\secref{mosag:spiralacq}) was never used during normal WEAVE
operations.

Of the 592 attempted LIFU OBs, 568 were successfully acquired and are analysed
in this section. The remaining 24 attempts failed at the acquisition stage
(primarily in sparse fields or under cloudy conditions), yielding an overall
LIFU acquisition success rate of 96 per cent. Eighteen of these failed attempts
occurred in early-2024, prior to the full implementation of the mitigation
strategies described in \secref{lifuacq:failure}. Failed acquisitions are
detected early in the OB and incur minimal observing overhead.

For all successful acquisitions, we extracted the total acquisition time,
measured from the first AG exposure to the moment acquisition was deemed
successful because the errors had fallen beyond the specified thresholds. In the
case of LIFU, the median acquisition time was 53 s, with 90 per cent completing
within 128 s. In the case of MOS, the distribution was bimodal: when guide stars
were visible inside the fibres from the very beginning and deploying the FPI was
not necessary, the median acquisition time was 76 s, with 90 per cent completing
within 309 s; when the FPI had to be deployed, the median rose to 309 s, with 90
per cent completing within 756 s.

We also extracted the number of iterations required by the AG to complete the
acquisition, which is at least 2: an image to estimate the initial error,
followed by the application of TCS offsets and a subsequent image to confirm
that the field is in the correct position. The LIFU pattern-recognition
algorithm converged in 2 iterations in 51 per cent of the cases, with 90 per
cent completing in at most 6 iterations. For the MOS acquisitions, the median
number of iterations was 7, with 90 per cent completing in at most 13
iterations.

Overall, unless no guide stars are visible in the MOS guide fibres (which is an
issue that occurs occasionally and is currently under investigation, with
promising developments over the last few months), automatic WEAVE acquisition is
expected to complete successfully within around a minute in the vast majority of
cases.

A direct comparison of acquisition performance between observing modes is given
in \tabref{tab:ag_summary}, including the dominant acquisition failure modes
encountered in routine operations, together with the corresponding fallback
strategies adopted for each observing mode.

\subsection{Guiding stability and accuracy}\label{results:guiding}
\subsubsection{Guiding corrections as a function of time}\label{results:guidingtimeseries}

In order to quantify the guiding stability, we present in \figref{fig:guidefull}
an ensemble of all autoguider corrections as a function of time for all OBs,
together with the median, $1\sigma$ (encompassing $\sim 68$ per cent of
corrections), and $2\sigma$ (encompassing $\sim 95$ per cent of corrections)
envelopes. Each exposure contributes a time series of centroid-offset
corrections with $t=0$ being the start of the exposure. Since there are hundreds
of lines in each plot, we adopt alpha-blending to avoid over-plotting and to
convey the statistical density of the trajectories. In this approach, each
individual time-series curve is drawn with a low opacity, so that regions where
many curves overlap appear progressively darker, while sparsely occupied regions
remain faint. This effectively transforms the stack of guiding-correction tracks
into a continuous density field, allowing the dominant trends and the spread of
the corrections to emerge visually without obscuring the underlying
distribution. The median and $\sigma$ envelopes are then overlaid with full
opacity for clarity, providing a clean statistical summary atop the blended
ensemble.

For LIFU (first row), where we do not apply corrections in rotation while
guiding, we only show the plots for $x$ and $y$ (the LIFU plate coordinates).
For the MOS AG (second and third rows) we plot the guiding corrections as a
function of time for instrument $x$ and $y$ (the MOS plate coordinates) and
rotator axes.

Despite the very large number of over-plotted sequences (149 for MOS and 568 for
LIFU), both figures reveal a strikingly narrow and stable locus. For both LIFU
and MOS, 68 per cent of the guiding corrections remain confined within $\pm 0.2$
arcsec over the entire OB, while 95 per cent stay within $\pm 0.4$ arcsec for
LIFU and $\pm 0.3$ arcsec for MOS. In rotation, 68 per cent of the corrections
are within $\pm 1$ millideg, while 95 per cent are smaller than $\pm 2$
millideg.

\begin{figure}
  \centering
  \includegraphics[width=\columnwidth]{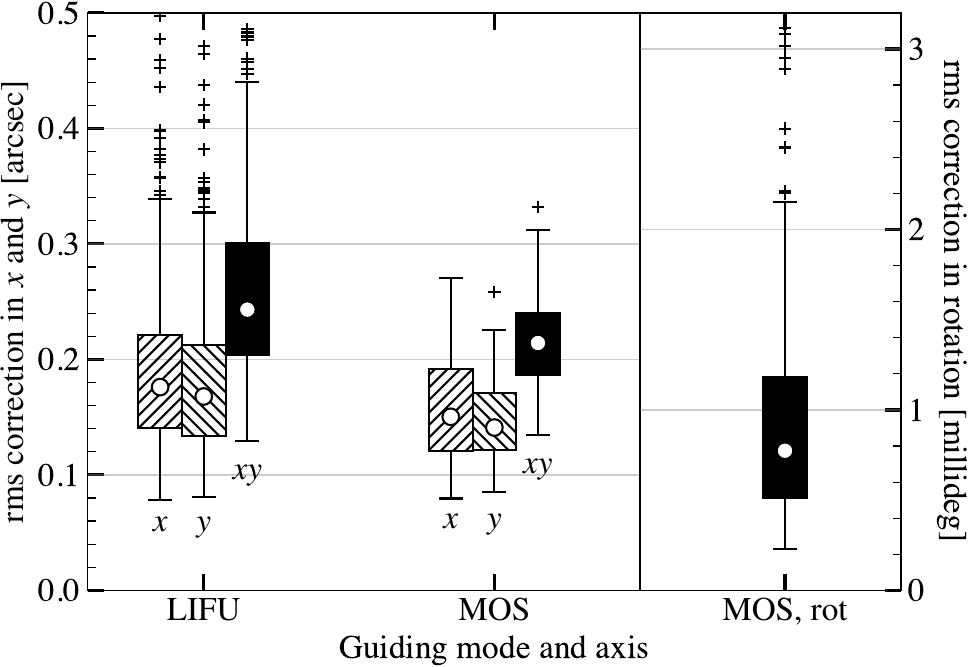}
  \caption{Box plots of the rms guiding residuals for all LIFU ($x$, $y$, $xy$)
  and MOS ($x$, $y$, $xy$ and rotation) OBs, calculated from the rms values of
  all of the lines plotted in \figref{fig:guidefull}. The median rms is marked
  with a white circle, while the outliers are marked with crosses. The bulk of
  exposures cluster between $0.2$--$0.3$~arcsec, indicating stable guiding
  throughout the observations.}
  \label{fig:rmsplot}
\end{figure}

For all $x$ and $y$ corrections, the widths of the distributions exhibit no
detectable growth with time, up to the typical duration of a WEAVE science OB of
almost one hour. This lack of broadening demonstrates that the control loop
remains stable throughout the exposure, with no evidence for systematic drift or
loop instability. This is not the case for the corrections in rotation, where we
see a slight increase in the $2\sigma$ envelope (though not in the $1\sigma$) at
late times (during the third exposure, approaching an hour of guiding). MOS OBs
are configured for a specific time and zenith distance, and the positioning of
the fibres becomes increasingly worse due to differential refraction (see
\appref{app:diffrefr}) as one departs from the configuration time, especially
for fibres at the edge of the plate. Although the AG does include differential
refraction of the calculation, the centroiding becomes less reliable as stars
shift towards the edge of the fibres, and it is precisely the stars that are
furthest away from the centre that determine the correction in rotation.
Nevertheless, an increase of less than 1 millideg does not have an appreciable
effect on the guiding quality, as it only corresponds to a shift of $\sim 30$
mas at the edge of the field.

Overall, the flat, time-independent envelopes provide a global validation of AG
performance across all observing conditions represented in the dataset, and
confirm that guiding jitter remains confined throughout the longest of the
exposures.

\begin{figure}
  \includegraphics[width=\columnwidth]{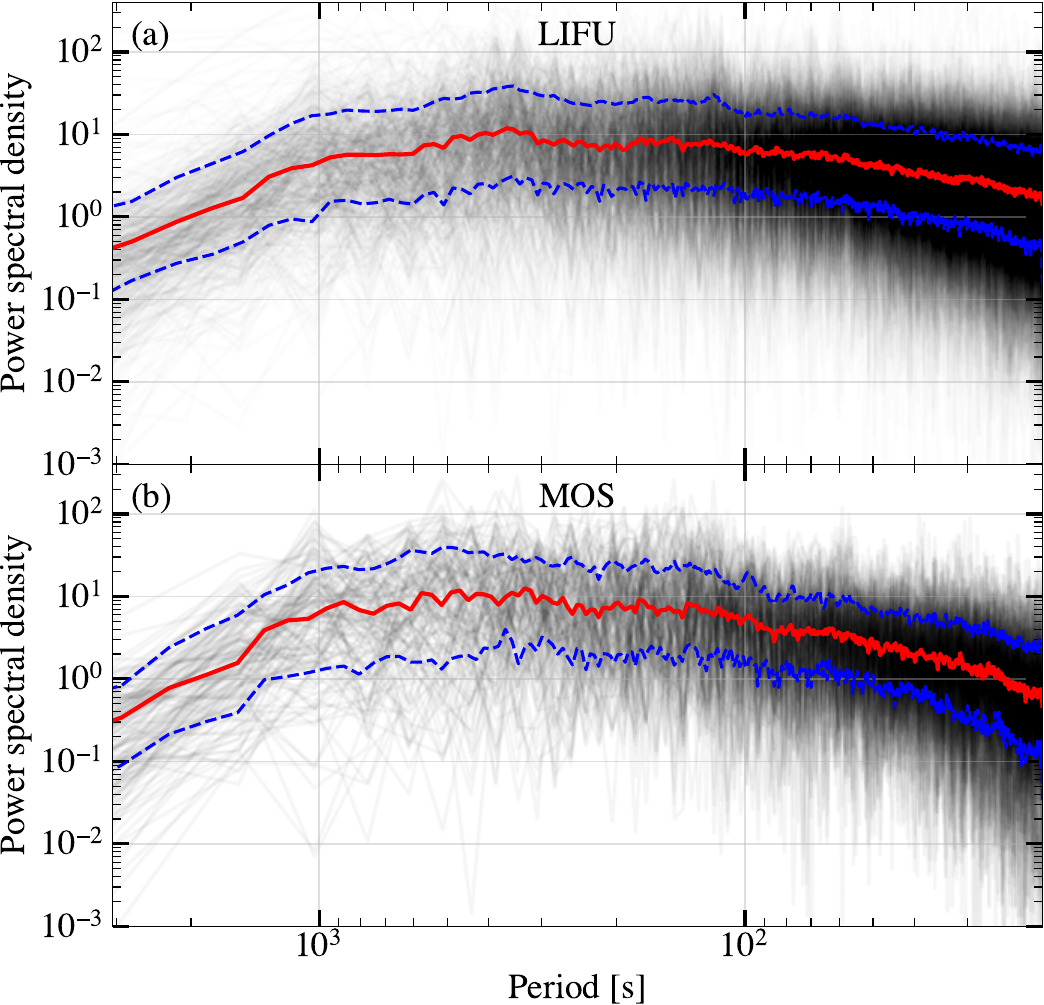}
  \caption{Power spectral density (PSD) of guiding corrections in $y$ as a
  function of period for (a) LIFU and (b) MOS OBs. Each grey curve shows
  the PSD of an individual time series of autoguiding corrections, after
  resampling the data on to a uniform time grid with 10-s spacing and computing
  a discrete Fourier transform. We also show the median values (solid red lines)
  and the $1\sigma$ envelope marking the $\sim 16$ and $\sim 84$ percentiles
  (dashed blue lines). The PSDs for the corrections in $x$ and (in the case of
  MOS) in rotation closely match the $y$-axis results, and are therefore omitted
  for brevity.}
  \label{fig:fourier}
\end{figure}

\begin{figure*}
  \centering
  \includegraphics[width=\textwidth]{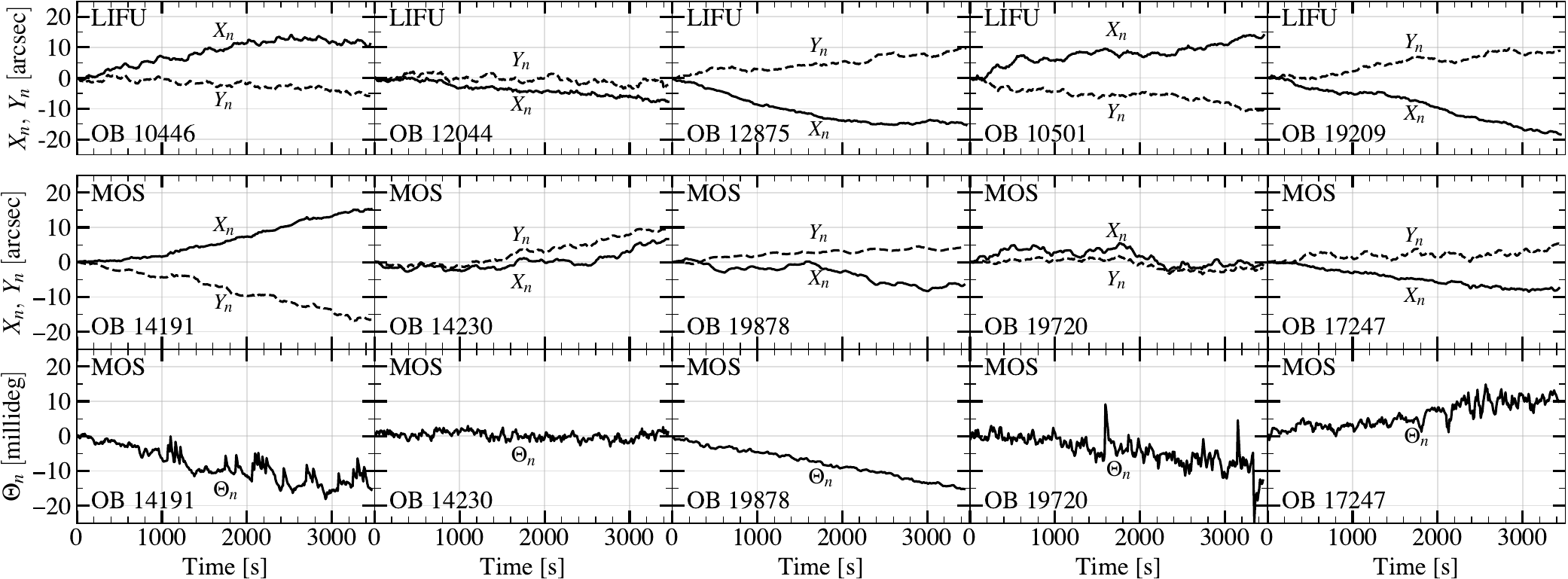}
  \caption{Cumulative guiding corrections as a function of time. The first row
  shows $X_n$ (solid) and $Y_n$ (dashed), as defined by \eqreft{eq:cumulative},
  for five LIFU OBs. The second and third rows show $X_n$, $Y_n$, and $\Theta_n$
  for five MOS OBs. In contrast to the guiding residuals, which remain centred
  around zero (\figref{fig:guidefull}), the cumulative corrections exhibit
  secular evolution over the course of an OB, reflecting the telescope and
  rotator drift compensated by the guiding loop.}
  \label{fig:cumulative_corrections}
\end{figure*}

\subsubsection{Root-mean-square guiding residuals}\label{results:guidingrms}
We also examined the rms guiding residuals during entire OBs, by applying
\eqreft{eq:rms} to each line in \figref{fig:guidefull} for calculating the rms
in $x$, $y$ and (in the case of MOS) rotation, and then \eqreft{eq:rmsxy} for
calculating the total $xy$ rms. Whereas the time-series plots highlight the
instantaneous behaviour of the guiding loop and its temporal stability within
individual exposures, the rms metric captures the integrated guiding performance
over the entire one-hour observation, and is therefore directly linked to the
reconstructed image quality delivered by the spectrograph. The results are shown
in \figref{fig:rmsplot}, where the boxes represent the interquartile range
(IQR), or the middle 50 per cent of the data, and the white circles inside the
boxes mark the median values; the whiskers extend to the usual $\pm 1.5$~IQR,
and the outliers are marked with open circles.

The plots show that the overwhelming majority of observations, independent of
the mode, exhibit rms values well within the 0.3-arcsec rms requirement. Median
and 90\textsuperscript{th}-percentile rms residuals for LIFU and MOS/mIFU
guiding, together with the corresponding cadences and operational statistics,
are summarized in \tabref{tab:ag_summary}.

We note that the rms corrections for MOS are not only overall smaller than for
LIFU, but they also exhibit far fewer outliers. This is because in MOS mode,
guiding uses the centroids of several guide fibres distributed across the focal
plane; as these fibres sample the field symmetrically, any residual distortions
(flexure, plate-scale anisotropy, or optical aberrations) tend to average out.
At the same time, using multiple centroids tends to cancel out centroiding
errors and noise, significantly lowering the overall rms and the spread of the
residuals compared to those obtained in the LIFU mode. Finally, MOS is almost
exclusively used when the seeing is below 1.3 arcsec, above which LIFU OBs are
generally observed. The fact that a small fraction of the LIFU OBs are observed
under worse conditions (DIMM seeing above 2 arcsec, bright moon, etc.) likely
contributes to these outliers.

\subsubsection{Power spectral analysis}\label{results:spectral}
To search for any periodic or quasi-periodic behaviour in the autoguiding
corrections, we computed the power spectral density (PSD) of the time series of
guiding residuals for both MOS and LIFU modes. The raw AG samples are
irregularly spaced in time, so for each exposure we first resampled the
corrections onto a uniform grid with 10~s spacing. A discrete Fourier transform
was then applied to the interpolated series to obtain the PSD as given by
\eqreft{eq:psd}, which we subsequently expressed as a function of period rather
than frequency for ease of interpretation (\figref{fig:fourier}).

Mechanical or control-loop oscillations within the telescope, rotator, or
autoguiding system would manifest themselves as narrow features or localised
power excesses at characteristic periods in the PSD. Identifying such signatures
would offer a direct means of diagnosing potential sources of guiding
instability.

The resulting spectra show no such behaviour. Instead, they are relatively
smooth and broadband over the range of periods (20 to 3000 s), with no evidence
of narrow peaks or enhanced power at specific frequencies. This indicates that
neither the telescope nor the autoguiding loop exhibits coherent oscillatory
behaviour within the accessible period range, and that the observed guiding
fluctuations are dominated by broadband stochastic processes (such as seeing and
centroiding noise) rather than by any coherent, systematic mechanical
oscillation or resonant motion. The similarity of the MOS and LIFU spectra
further suggests that the absence of narrow spectral features reflects a
property of the telescope--atmosphere system rather than of a specific observing
mode.

\subsubsection{Analysis of cumulative corrections}\label{results:cumulative}
So far we have focused on the guiding residuals, defined as the measured
centroid offsets in detector coordinates ($x$, $y$, and rotation) after each
guiding update. These characterize the stability of the closed-loop system and
demonstrate that the AG maintains the required accuracy throughout typical WEAVE
observations. However, in the presence of a PID control loop
(\secref{intro:agdesign}), long time-scale telescope and rotator drift are
largely removed from the residuals and are instead encoded in the accumulated
guiding corrections.

\begin{figure*}
  \centering
  \includegraphics[width=\textwidth]{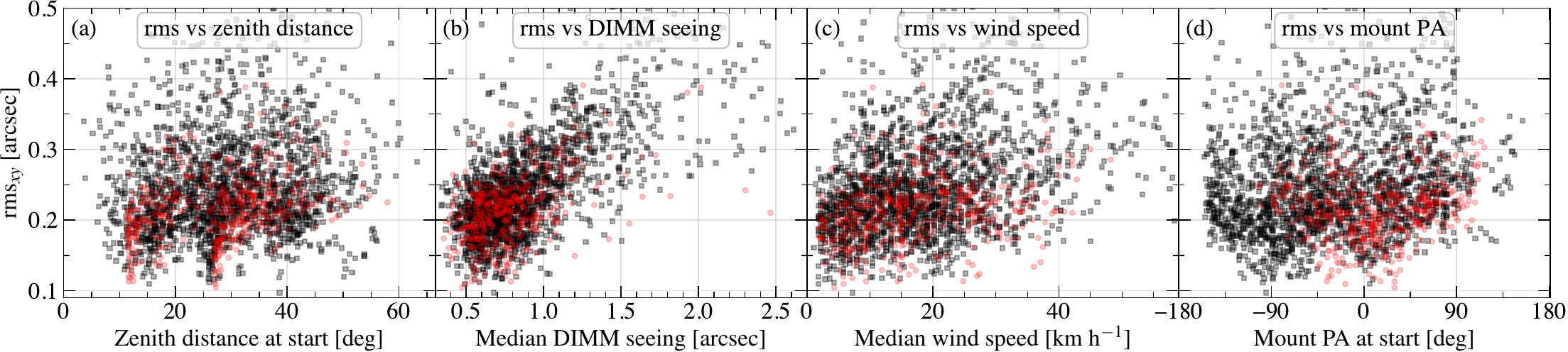}
  \caption{Dependence of the rms guiding residuals on various observing
  conditions retrieved from the science FITS headers:
  (a) zenith distance at the start of the exposure,
  (b) median DIMM seeing during the exposure,
  (c) median wind speed, and
  (d) mount position angle at the start of the exposure.
  Data points are shown with black squares for LIFU and red circles for MOS.
  Both seeing and wind speed exhibit positive correlations with rms
  ($R^2\approx 0.22$ and $R^2\approx 0.09$, respectively), but the large scatter
  indicates that other factors (such as guide star SNR and wind direction)
  modulate the instantaneous guiding performance.}
  \label{fig:conditions}
\end{figure*}

In the WEAVE guiding implementation, the measured offsets are transmitted
directly to the telescope control system and integrated with a fixed gain of
$0.3$. This choice is rooted in control theory (e.g.,
\citealp{parr1996,jenkins1998}), as servo gains in the range 0.2--0.5 suppress
overshoot while maintaining effective correction of low-frequency tracking
errors. The exact value of 0.3 has been determined empirically for the WHT since
the early 1990s and has consistently yielded stable, non-oscillatory guiding
performance across instruments; we therefore adopted the same value for WEAVE.
As a result of the fixed gain, the cumulative sum of the measured offsets
(\eqrefp{eq:cumulative}) provides a reconstruction of the total corrections
applied over the course of an OB. 

\figref{fig:cumulative_corrections} shows several examples of cumulative
corrections as a function of time for several OBs (which we selected for having
particularly large drifts). In contrast to the residuals, which fluctuate around 
zero with no measurable long-term trend (e.g., \figref{fig:guidefull}), the
cumulative corrections often exhibit clear secular evolution over the duration
of an exposure, reaching amplitudes of up to a few tens of arcsec in translation
and rotation over a typical 1-h OB. The sign and magnitude of these trends vary
between OBs, reflecting their dependence on sky position and rotator angle.

The presence of such trends indicates that the guiding loop is actively
compensating for low-frequency telescope and rotator motion. Importantly, these
cumulative corrections are built up through many small, incremental adjustments
at each guiding step, consistent with the observed sub-arcsec residuals at all
times. The large cumulative amplitudes therefore do not imply large
instantaneous pointing errors, but rather the gradual cancellation of slow,
continuous drift.

Taken together, the residual and cumulative analyses provide a consistent
picture: the AG maintains tight control of the pointing and rotation,
suppressing residual errors to well below the instrument requirements, while
continuously compensating for underlying telescope and rotator motion on
time-scales comparable to the duration of a WEAVE OB.

\subsection{Robustness across observing conditions}\label{results:rmsconditions}
To assess whether WEAVE’s guiding performance varies systematically with
observing conditions, we examined the dependence of the per-exposure total rms
guiding residuals (\eqrefp{eq:rmsxy}) on a range of environmental and telescope
parameters. For each exposure we extracted the zenith distance, airmass,
instrument rotator angle, wind speed, DIMM seeing, and ambient temperature from
the FITS headers, and correlated these with the measured rms values. Across the
full dataset, we found no significant correlation between rms guiding stability
and most of these parameters (see \figref{fig:conditions}). In particular,
neither zenith distance nor airmass shows a measurable trend, indicating that
flexure, refraction-induced centroid shifts, or any other distortions are
effectively absorbed by the guiding loop. Likewise, the rotator angle exhibits
no impact on the rms levels, demonstrating that the guiding system is
insensitive to rotator orientation.

We only find a weak but physically plausible dependence of the per-exposure rms
guiding residuals on atmospheric seeing and on local wind speed. Poorer DIMM
seeing increases the instantaneous centroid wander of the guide star because
atmospheric turbulence produces faster and larger image motion and blurring of
the point-spread function; this raises the centroid measurement noise and
therefore the closed-loop residuals. Wind acts through two related mechanisms:
turbulent air inside the dome and along the optical path (which degrades
effective seeing locally), and direct mechanical excitation of the telescope
structure (windshake), which introduces real image motion on the time-scales
sampled by the autoguider.

Despite being statistically detectable, these dependencies are modest in
magnitude (explaining only $\approx 22$~per cent and $\approx 9$~per cent of the
variance for DIMM seeing and wind, respectively, as measured by $R^2$). The
large scatter in both plots reveals that while seeing and wind are real
contributors to guiding performance, they are not the dominant drivers for any
single exposure. We also note that the DIMM seeing plot should be read in the
context of the year-round median seeing at the WHT, which is 0.69 arcsec
(\citealp*{munoz1997}; \citealp{wilson1999}).

Overall, this analysis demonstrates that WEAVE's autoguiding system delivers
consistent and robust performance across the full range of operational
conditions encountered at the WHT, with only a mild, physically expected
dependence on atmospheric seeing.


\section{Discussion}\label{sec:discussion}
The last two years of commissioning and early survey operations have produced a
rich dataset that allows a comprehensive assessment of the AG's behaviour across
the full range of observing conditions at the WHT. The results
(\secref{sec:results}) demonstrate that both guiding modes meet or exceed the
top-level requirements, with performance that is robust to variations in seeing,
weather conditions, and telescope pointing. They also prove that the AG is
highly automatized, which aligns with WEAVE's queue-scheduled operational model.
In this section we compare the performance to the design requirements, examine
the consequences for science operations, summarize the lessons learned, and
outline upcoming improvements.

\subsection{Comparison with design requirements}\label{discussion:vsdesign}
The WEAVE AG was designed to guarantee 0.3-arcsec rms placement and
long-duration stability of targets, both across the 2\textdegree\ field in
MOS/mIFU modes, and on to the LIFU central fibre, while supporting continuous,
unattended operation within the OCS sequencer environment. The results obtained
to date confirm compliance with all essential requirements.

Both acquisition modes achieve the necessary speed and reliability
(\secref{results:acquisition}). In LIFU mode, the pattern-recognition-based
acquisition procedure converges rapidly (median time under 1 min) and succeeds
across sparse, crowded, and non-uniform fields, including under thin clouds. MOS
acquisition typically converges within a median time of under 2 min, even in
cases where guide stars are at or slightly beyond the edges of the fibres. The
principal fallback mechanism (FPI-assisted acquisition) has proven robust in
occasional cases of large initial offsets, and in practice the required
acquisition time is dominated by the quality of the initial blind pointing
rather than by limitations in the AG itself.

Both systems deliver stable guiding over the full 1-h duration of a typical
WEAVE OB. The time-series analysis (\secref{results:guidingtimeseries}) shows
that the guiding corrections remain well-behaved, with no secular drift or
increase in magnitude. The AG corrects both short-timescale guiding errors and
long-term pointing drifts, as confirmed by the analysis of cumulative
corrections (\secref{results:cumulative}), which shows that slow systematic
offsets are effectively removed during closed-loop operation. Drifts that would
be present due to atmospheric differential refraction are prevented by per-frame
astrometric recalculations in MOS (recently implemented in LIFU as well, see
below).

The measured median guiding rms lies around 0.2 arcsec
(\secref{results:guidingrms}), comfortably meeting the 0.3-arcsec requirement
and demonstrating that the TCS--AG control loop performs as designed. The
dependence of the guiding rms on external conditions is mild
(\secref{results:rmsconditions}). The absence of strong degradation under poorer
seeing or increased wind confirms that the AG measurements remain photon-limited
in most cases, with only a weak correlation with DIMM seeing ($R^2\approx 0.22$)
and wind speed ($R^2\approx 0.09$). This stability is especially important when
the scheduler must continue executing programmes as conditions evolve over the
course of a night.

Finally, the broadband power-spectral density (\secref{results:spectral}) shows
no coherent oscillatory features, confirming that neither mechanical resonances
nor algorithmic instabilities introduce periodic structure into the guiding
residuals that could otherwise propagate into the science data.

Overall, the measured performance demonstrates that the system satisfies the
guiding accuracy, stability, and automation criteria set forth in the WEAVE
top-level requirements.

\subsection{Impact on science performance}\label{discussion:impactonscience}
The primary function of the AG is to ensure accurate positioning of science
targets in the WEAVE focal plane (whether on to MOS fibres, the mini-IFUs, or
the central LIFU fibre) and to maintain that placement throughout exposures.
The guiding behaviour therefore sets the ceiling on fibre-coupling efficiency
and on the spectroscopic throughput delivered to survey programmes.

The typical MOS guiding rms of $\sigma=0.21$ arcsec (\tabref{tab:ag_summary})
implies only minor flux losses for the 1.3-arcsec diameter science fibres
(\citealp{mignot2020}). The distribution of offsets, $R$, over the complete set
of fibres is governed by a Rayleigh distribution, such that the proportion of
fibres out of position by more than $R$ arcsec is given by
\begin{equation}
C(>R) = \exp(-R^2/\sigma^2).
\end{equation}
This would correspond to offsets of $\approx 0.17$ and 0.32 arcsec for the
median and 90\textsuperscript{th} percentiles, respectively. Assuming 0.7~arcsec
seeing and a Moffat PSF with $\beta = 2.5$, a perfectly centred fibre captures
0.672 of the incident flux. Offsets corresponding to the median and
90\textsuperscript{th} percentiles would therefore reduce this to 0.647 (96 per
cent) and 0.582 (87 per cent), respectively. Thus, even at the
90\textsuperscript{th} percentile, flux losses remain under $\sim 13$ per cent.
These estimates ignore fibre-to-fibre throughput variations, which are still
under investigation.
Because the quoted rms reflects the combined behaviour of the POS--AG--TCS
system, the contribution from the AG alone is even smaller. As a result, the AG
does not impose any additional throughput limit on survey operations: fields
with well-positioned fibres can be expected to deliver science-grade throughput
irrespective of the distribution or brightness of the guide stars.

In both mIFU and LIFU observations, stable guiding ensures that the targets
remain fixed with respect to their IFU footprints throughout an exposure. This
preserves a stable sky-to-IFU mapping, ensuring that wavelength-dependent PSF
variations reflect atmospheric and optical effects rather than being distorted
by time-dependent guiding jitter. The absence of coherent oscillatory components
in the guiding corrections further ensures that the PSF does not acquire
time-dependent structure, which could otherwise introduce systematic biases in
the kinematic and dynamical measurements. In addition, for LIFU, the placement
of the target on the central fibre is completely determined by the accuracy of
the acquisition algorithm. 

The successful LIFU acquisition and guiding of moving targets
(\secref{lifuag:nonsiderealguiding}), as demonstrated for 3I/ATLAS, expands the
range of scientifically viable programmes to include Solar System and
interstellar objects, where positional accuracy and uninterrupted tracking are
critical.

Finally, the high AG uptime and streamlined, automated operation have enabled
WEAVE surveys to proceed with minimal human intervention. This has reduced
operational overheads, increased observing efficiency, and allowed the OCS to
execute long, mixed-programme queues reliably.

\subsection{Lessons learned from first years of operations}\label{discussion:lessonslearned}
Routine use of the AG has provided us with several insights that shaped
subsequent improvements.

\begin{description}
\item \textit{Value of a comprehensive simulation mode.}
A key lesson learned from the WEAVE AG development is the value of a
high-fidelity simulation mode (\secref{sec:sim}) that extends beyond engineering
use. Providing astronomers and telescope operators with access to a realistic,
end-to-end simulated OCS and AG environment well before commissioning enabled
early training, reduced operational risk, and encouraged constructive feedback
on interfaces and workflows. This significantly lowered the cognitive load
during commissioning, when attention could be focused on science tasks rather
than software familiarity. In addition, the simulation environment has
repeatedly demonstrated its importance not only during early development, but
throughout routine operations. The ability to replay on-sky problems
deterministically and test new algorithms off-sky accelerated debugging and
reduced risk to night-time observing. We strongly recommend that future complex
survey instruments invest in simulation environments that are suitable for both
software testing and for long-term user training and cross-team integration.
\item \textit{The critical importance of accurate and automated calibrations.}
Frequent recalibration of fibre positions and orientations
(\secref{mosag:calibs}) has proven essential, as even small mechanical shifts in
the optical system translate into measurable centroid errors. Automation of
these calibrations was therefore a necessary operational enabler for nightly
reliability.
\item \textit{Rotational degeneracy in off-axis guiders.}
The intrinsic geometric degeneracy in the LIFU AG rotation solution
(\secref{lifuacq:failure}) proved more restrictive in practice than initially
expected. While acquisition is robust, applying rotation corrections during
guiding without filtering introduces noise-driven oscillations. This emphasizes
the need for additional filtering, or for conditioning metrics based on actual
centroid error estimates, resulting in a more conservative application of
rotation updates in off-axis systems.
\item \textit{Rotational acquisition as a diagnostic for rotator encoder health.}
A useful side effect of LIFU rotational acquisition is the ability to identify
anomalies in telescope rotator behaviour during normal observing operations.
Because acquisition explicitly solves for the required field rotation, unusually
large initial corrections (over 1\textdegree, rather than the typical
0.1\textdegree) reveal inconsistencies between demanded and measured rotator
position. Several such events observed during commissioning were traced to
rotator encoder glitches requiring maintenance. While rotator performance is
routinely monitored through dedicated engineering procedures, LIFU acquisition
provides an independent on-sky verification of system health.
\item \textit{Sensitivity to mechanical fibre-placement errors.}
In MOS mode, the rare cases of mispositioned fibres (\secref{mosag:autoacq})
confirm that mechanical placement accuracy sets the limiting floor for a small
fraction of fields. The AG algorithms function correctly, but systematic
centroid bias from off-centre stars limits the attainable guiding rms. These
events motivated improvements in fibre-positioning diagnostics and highlighted
the need for continuous monitoring and early flagging of positioning errors.
\item \textit{Interaction of seeing, focus, and fibre geometry.}
The autofocus results (\secref{mosag:multifocus}) show that fibre-edge effects
and variable seeing lead to non-ideal behaviour in wide focus sweeps. Awareness
of these limitations is essential for correct interpretation of focus curves,
and has driven improvements to the focusing workflow.
\end{description}

Collectively, these experiences highlight that long-term, automated operation
depends as much on robust software and calibration strategies as on the
underlying algorithms.

\subsection{Planned improvements and future development}\label{discussion:future}
The consolidated performance metrics presented in \tabref{tab:ag_summary}
demonstrate that the AG system meets its design requirements across all
observing modes, while highlighting the distinct operational characteristics of
LIFU and MOS/mIFU guiding.

Our analysis demonstrates that the WEAVE autoguider already delivers highly
stable guiding over long science exposures, but also points to several avenues
for future improvement. A number of enhancements are already in progress or
planned for the near future:

\begin{description}
\item \textit{Multi-star LIFU guiding with continuous differential refraction
updates.} The most impactful enhancement is the introduction of multi-star
guiding in LIFU mode, which is expected to reduce the measured rms to similar
levels as in MOS (\figref{fig:rmsplot}). This development is already in
progress: preliminary tests conducted in December 2025 demonstrate that
multi-star guiding performs reliably, although additional data are required to
quantify the improvement over single-star guiding. At the same time, we have
extended the MOS-style per-frame astrometric recalculation to LIFU, which
eliminates any drift induced by changes in refraction during long exposures.
Across the FOV of the LIFU AG (3.8 arcmin), however, differential refraction is
negligible except at the largest zenith distances used in WEAVE OBs
($z=60$\textdegree, where it reaches $\sim 0.23$~arcsec, or $\sim 1$~px).
\item \textit{LIFU guiding in rotation.} Once multi-star guiding was available,
extending the control loop to include rotation proved technically
straightforward. However, the effectiveness of rotational corrections is limited
by the degeneracy described in \secref{lifuacq:failure}. We are currently
implementing Kalman-filtered rotation updates that preserve long time-scale
rotational drift corrections while suppressing frame-to-frame noise. Depending
on the on-sky performance of these filters, further conditioning metrics may
also be needed to disable rotation updates automatically when the solution
becomes ill-posed.
\item \textit{Improved acquisition in sparse fields.} In sparse fields (or
cloudy conditions), where the current LIFU acquisition algorithm occasionally
fails, we plan to explore triangle-based matching algorithms (e.g.,
\citealp{groth1986,valdes1995,pal2006}; \citealp*{beroiz2020}). The extra
computational cost would be of little relevance in sparse fields, but such an
approach would have the advantage of only requiring three sources to yield a
match. As a last-resort fallback, we are also investigating point-pattern
matching techniques that dispense with explicit asterism construction
(\citealp{murtagh1992}).
\item \textit{Full support for non-sidereal observations.} Following the
successful demonstration with comet 3I/ATLAS, additional features will be
introduced to support variable-rate ephemerides, improved metadata handling, and
coordinated scheduling for time-critical Solar System campaigns.
\item \textit{Mechanical improvements.} Another priority is addressing the known
cases of suboptimal fibre positioning (described in \secref{mosag:autoacq}),
which occasionally limits the number of usable guide stars or leads to
asymmetric centroiding performance; improving the mechanical accuracy and
repeatability of fibre placement would further stabilise the guiding loop.Work
is also underway to make the MOS/mIFU guide fibres parfocal
(\secref{sec:internalfocus}), which is expected to lower the intrinsic
fibre-to-fibre PSF variation and improve the reliability of autofocus
procedures.
\item \textit{Further development of the simulation environment.}
Extensions to the simulated environment are planned to include extended sources,
PSF variability across the field, and non-sidereal motion models, allowing 
more comprehensive end-to-end testing of new AG features.
\end{description}

Together, these developments will further strengthen the accuracy and
reliability of the WEAVE AG, ensuring that the delivered data fully meet the 
survey's scientific goals as it enters full operational cadence.


\section*{Acknowledgements}
We express our gratitude to the reviewers of the paper, Steven Beard and Will
Sutherland, whose thorough and insightful comments significantly improved the
quality of the manuscript.
We thank Scott Trager, Johan Pragt, and Rik ter Horst for permission to
reproduce a modified version of the LIFU fibre head schematic (inset (a) of
\figref{fig:agdiagram}) originally developed at ASTRON.
We also thank Mike Irwin for his valuable input on the flux-loss calculations
presented in \secref{discussion:impactonscience}.

Funding for the WEAVE facility has been provided by UKRI--STFC, the University
of Oxford, NOVA, NWO, Instituto de Astrofísica de Canarias (IAC), the Isaac
Newton Group partners (STFC, NWO, and Spain, led by the IAC), INAF, CNRS--INSU,
the Observatoire de Paris, Région Île-de-France, CONACYT through INAOE, the
Ministry of Education, Science and Sports of the Republic of Lithuania, Konkoly
Observatory (CSFK), Max-Planck-Institut für Astronomie (MPIA Heidelberg), Lund
University, the Leibniz Institute for Astrophysics Potsdam (AIP), the Swedish
Research Council, the European Commission, and the University of Pennsylvania.
The WEAVE Survey Consortium consists of the ING, its three partners, represented
by UKRI STFC, NWO, and the IAC, NOVA, INAF, CNRS--INSU, INAOE, Vilnius
University, FTMC--Center for Physical Sciences and Technology (Vilnius), and
individual WEAVE Participants. Please see the relevant footnotes for the WEAVE
website\footnote{\href{https://weave-project.atlassian.net/wiki/display/WEAVE}{https://weave-project.atlassian.net/wiki/display/WEAVE}} 
and for the full list of granting agencies and grants supporting 
WEAVE\footnote{\href{https://weave-project.atlassian.net/wiki/display/WEAVE/WEAVE+Acknowledgements}{https://weave-project.atlassian.net/wiki/display/WEAVE/WEAVE+ Acknowledgements}}.

The software presented in this article makes extensive use of 
\textsc{slalib} (\citealp{wallace2014}),
\textsc{numpy} (\citealp{harris2020}), 
\textsc{astropy} (\citealp{astropy2022}),
\textsc{photutils} (\citealp{bradley2025}),
\textsc{scipy} (\citealp{virtanen2020}),
\textsc{ds9} (\citealp{joye2003}),
\textsc{pyds9}\footnote{\href{https://github.com/ericmandel/pyds9}{https://github.com/ericmandel/pyds9}},
\textsc{wxPython}\footnote{\href{https://wxpython.org/}{https://wxpython.org/}},
\textsc{matplotlib} (\citealp{hunter2007}), and
Redis\footnote{\href{https://redis.io/}{https://redis.io/}}.
It also uses \textsc{vizier}\footnote{\href{https://vizier.cds.unistra.fr/}{https://vizier.cds.unistra.fr/}} (\citealp*{ochsenbein2000}),
\textsc{hips2fits}\footnote{\href{https://alasky.cds.unistra.fr/hips-image-services/hips2fits}{https://alasky.cds.unistra.fr/hips-image-services/hips2fits}},
and \textsc{aladin}\footnote{\href{https://aladin.cds.unistra.fr/}{https://aladin.cds.unistra.fr/}} (\citealp{bonnarel2000}),
services provided by CDS, Strasbourg Observatory, France.

This work has also made use of data from the European Space Agency (ESA) mission
Gaia\footnote{\href{https://www.cosmos.esa.int/gaia}{https://www.cosmos.esa.int/gaia}} (\citealp{gaia2023}),
processed by the Gaia Data Processing and Analysis Consortium
(DPAC)\footnote{\href{https://www.cosmos.esa.int/web/gaia/dpac/consortium}{https://www.cosmos.esa.int/web/gaia/dpac/consortium}}. 
Funding for the DPAC has been provided by national institutions, in particular
the institutions participating in the Gaia Multilateral Agreement.

\section*{Conflict of interest}
The authors declare no conflict of interest.

\section*{Data availability}
The data underlying this article (e.g., measurements used for plots) will be
shared on reasonable request to the first author.
The simulated camera described in \secref{sim:agcamera} is released under an MIT
license on GitHub, see
\href{https://github.com/egafton/siren}{https://github.com/egafton/siren}.


{
 \hypersetup{urlcolor=bibref} 
 \bibliographystyle{rasti} 
 \bibliography{agpaper}
}


\appendix
\section{Differential refraction}\label{app:diffrefr}
The effect of the atmosphere on an incident light ray is to cause a deviation
such that the `apparent' zenith distance measured at ground level is less than
the `true' zenith distance that would have been measured in the absence of an
atmosphere. Differential refraction takes place along the great circle passing
through the target and the zenith, i.e., it is perpendicular to the horizon.
One can decompose this effect into
(1) a spectral, chromatic differential refraction, due to the fact that incident
light rays of different wavelengths $\lambda$ at the same zenith distance are
refracted by a different amount, according to the index of refraction
$n(\lambda)$; in practice, an ADC is normally used to compensate for the
spectral differential refraction, and since the WEAVE PFC is equipped with one,
this correction is of little concern to the WEAVE AG;
(2) a spatial, achromatic differential refraction, due to the fact that the FOV
contains points at different zenith distances, for which the refraction will be
different; this is the differential refraction which the AG must account for.

\begin{figure}
  \centering
  \includegraphics[width=\columnwidth]{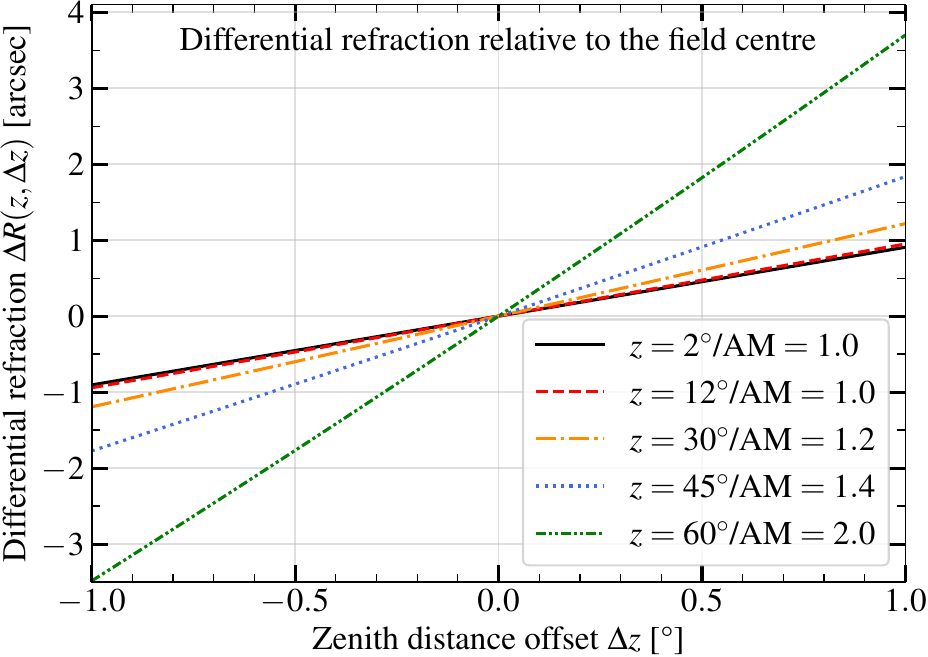}
  \caption{Typical field differential refraction values relative to the field
  centre, computed with the \textsc{slalib} function \texttt{slaREFRO}, across
  the WEAVE FOV of 2\textdegree. We used the mean weather conditions at the WHT
  (temperature 5.8~\textdegree{}C, pressure $770.6$~mbar, relative humidity
  $36.7$ per cent). The various lines represent observations at various zenith
  distances, ranging from 2\textdegree\ (where $\Delta R$ at the lower field
  edge is 0.9~arcsec) to 60\textdegree\ (where $\Delta R$ at the lower field
  edge is 3.7~arcsec); the corresponding airmass (AM) is also shown in the
  legend. Note that the plot grows increasingly asymmetric with $z$, as the
  differential refraction at the lower field edge (positive $\Delta z$) is
  larger than the corresponding $\Delta R(-\Delta z)$.}
  \label{fig:diffrefr}
\end{figure}

The accurate calculation of atmospheric refraction is onerous, requiring the
integration of the light path through a model atmosphere. In practice, assuming
the telescope axis is pointing at a zenith distance $z$, the refraction can be
approximated by (e.g., \citealp{woolard1966})
\begin{equation}\label{eq:R}
R(z) = A \tan z + B \tan^3 z,
\end{equation}
\noindent where the coefficients $A$ and $B$ account for the curvature of the
Earth, the structure of the atmosphere above the observing site, and the current
temperature, pressure and relative humidity (e.g., \citealp{saastamoinen1972});
this correction is taken into account by the TCS when pointing to a given field.
In addition, any point in the image that has a zenith distance offset by
$\Delta z$ from the zenith distance $z$ of the field centre will exhibit a
differential refraction with respect to the field centre. The magnitude of this
effect can be estimated through a Taylor expansion of \eqreft{eq:R} around $z$,
noting that the first-order expansion is antisymmetric in $\Delta z$, and that
inclusion of higher-order terms is necessary to capture the asymmetric increase
of refraction away from zenith (see \figref{fig:diffrefr}).

In practice, though, the WEAVE AG does not directly compute the refraction.
Instead, by providing the current weather conditions during the conversion from
apparent to observed coordinates using the \textsc{slalib} function
\texttt{slaAOP}, the observed coordinates automatically include the refraction
term $R$ as given by \eqreft{eq:R} for every star in the image, and for the
field centre. When projecting the observed coordinates on to the tangent plane
with the observed field centre as the tangent point, the resulting standard
coordinates $(\xi,\eta)$ will then be $(0,0)$ for the field centre, and will
only include the differential refraction for all of the other points,
essentially taking out the correction already performed by the TCS, and only
leaving in the correction that still needs to be applied by the AG.

\section{Coordinate systems in WEAVE}\label{app:coords}
The astrometry calculations that predict where a star should appear on the
detector involve the conversion between many types of coordinates, which we
describe here for reference.
\begin{description}
\item \textit{Mean RA, Dec} (rad) are the coordinates $\alpha_{\rm ICRS}$,
$\delta_{\rm ICRS}$ provided by the surveys (or retrieved from the Gaia
catalogue, in the case of the finding chart stars) together with the proper
motions $\mu_\alpha$ and $\mu_\delta$, epoch and equinox $J$, parallax $\pi$,
and radial velocity $R_{\rm v}$, and are always expressed in the ICRS frame.
\item \textit{Apparent RA, Dec} (rad) are the geocentric apparent coordinates
$\alpha_{\rm ap}$, $\delta_{\rm ap}$ with respect to the true equator and
equinox of date, and are computed from the mean RA, Dec by taking into account
precession--nutation, annual aberration, and light deflection (the gravitational
lens effects of the sun) for the current barycentric dynamical time ${\rm TDB}$,
using the \textsc{slalib} (\citealp{wallace2014}) function \texttt{slaMAP}:
\begin{equation}
\alpha_{\rm ap}, \delta_{\rm ap} = {\rm slaMAP}(\alpha_{\rm ICRS}, \delta_{\rm ICRS},
\mu_\alpha, \mu_\delta, \pi, R_{\rm v}, J, {\rm TDB}).
\end{equation}
\noindent Note that while the AG uses apparent coordinates as the intermediate
step between ICRS and observed place (which is the most natural choice when
using \textsc{slalib}), the positioner software uses CIRS as the intermediate
frame (the better choice with the
\textsc{sofa}\footnote{\href{http://www.iausofa.org/}{http://www.iausofa.org/}}
library, used by the positioner); the observed places calculated by the two
applications are otherwise identical to milliarcsec precision, mostly due to the
different precession--nutation models used by the two underlying libraries.
\item \textit{Observed RA, Dec} (rad) are the position $\alpha_{\rm ob}$,
$\delta_{\rm ob}$ as seen by a perfect theodolite at the mean longitude
$\lambda$ and geodetic latitude $\phi$ of the observer, and are computed from
the apparent RA, Dec  by taking into account the observer's current date and
time ${\rm UTC}$, location and height above sea level $H$, current polar motion
coordinates $x_{\rm p}$ and $y_{\rm p}$, current
${\rm DUT}\equiv{\rm UT1}-{\rm UTC}$ correction, and the atmospheric refraction
as a function of temperature $T$, pressure $p$, and relative humidity $\varphi$,
using the \textsc{slalib} function \texttt{slaAOP}:
\begin{equation}
\alpha_{\rm ob}, \delta_{\rm ob} = {\rm slaAOP}(\alpha_{\rm ap}, \delta_{\rm ap},
{\rm UTC}, {\rm DUT}, \lambda, \phi, H, x_{\rm p}, y_{\rm p}, T, p, \varphi).
\end{equation}
\item \textit{Standard coordinates} $\xi$, $\eta$ (rad) are the tangent plane
coordinates, computed through a gnomonic projection of the observed RA, Dec with
respect to observed coordinates $\alpha_{\rm ob,fc}$, $\delta_{\rm ob,fc}$ of
the field centre as the tangent point, with the $\eta$-axis pointing North and
the $\xi$-axis pointing East, using the \textsc{slalib} function
\texttt{slaDS2TP}:
\begin{equation}
\xi, \eta = {\rm slaDS2TP}(\alpha_{\rm ob}, \delta_{\rm ob}, \alpha_{\rm ob,fc}, \delta_{\rm ob,fc}).
\end{equation}
\item \textit{Focal plane coordinates} $x_{\rm fp}$, $y_{\rm fp}$ (mm) are
related to standard coordinates by the focal length of the telescope $f$, with
an additional fifth-order power-law correction $\mathcal{F}(r)$ for the radial
distortion:
\begin{equation}
\begin{bmatrix}x_{\rm fp} \\ y_{\rm fp}\end{bmatrix}
= f\mathcal{F}\left(\sqrt{\xi^2+\eta^2}\right)
\begin{bmatrix}\xi \\ \eta\end{bmatrix}.
\end{equation}
\item \textit{Plate coordinates} $x_{\rm pl}$, $y_{\rm pl}$ (mm) are obtained
from the focal plane coordinates by accounting for the translation ($x_0$,
$y_0$) and rotation ($r_0$) offsets between the plate and the focal plane, as
well as the sky PA.
\begin{equation}
\begin{bmatrix}x_{\rm pl} \\ y_{\rm pl}\end{bmatrix}
= \begin{bmatrix}\cos\left(r_0 + {\rm PA}\right) & -\sin\left(r_0 + {\rm PA}\right) \\
                 \sin\left(r_0 + {\rm PA}\right) & \phantom{-}\cos\left(r_0 + {\rm PA}\right)\end{bmatrix}
\begin{bmatrix}x_{\rm fp} \\ y_{\rm fp}\end{bmatrix} + 
\begin{bmatrix}x_0 \\ y_0\end{bmatrix}.
\end{equation}
In MOS mode, the origin of the plate coordinate system is defined by fiducial
marks etched onto each field plate (\citealp{dalton2020}). In LIFU mode, where
no configurable plate exists, the coordinate origin is instead defined such that
the central LIFU fibre lies at $(x_{\rm pl},y_{\rm pl})=(0, 20)$~mm. The
translation offset between the plate reference frame and the telescope optical
axis, as well as the rotational offset between the plate and rotator axes, have
been determined from on-sky calibration measurements; a detailed description
will be presented by G. Dalton et al. (in preparation).
\item \textit{Detector coordinates} $x_{\rm px}$, $y_{\rm px}$ (px) are related
(in the case of LIFU) to the plate coordinates by the pixel scale $s$ and offset
$\Delta x$, $\Delta y$ of the detector:
\begin{equation}
\begin{bmatrix}x_{\rm px} \\ y_{\rm px}\end{bmatrix}
= s\left(\begin{bmatrix}x_{\rm pl} \\ y_{\rm pl}\end{bmatrix} -
  \begin{bmatrix}\Delta x \\ \Delta y\end{bmatrix}\right).
\end{equation}
\noindent In the case of MOS, the transformation is done relative to the
position of each fibre on the CCD, while also taking the fibre orientation into
account.
\end{description}

\section{A linearized rigid-transformation solver}\label{app:transrot}
To determine the telescope offsets from measured and expected guide star
positions, the initial version of the AG system employed a full non-linear
least-squares fit of a rigid rotation and translation, with the corresponding
normal equations solved via singular value decomposition (SVD). This procedure
had previously been used at the WHT by the similar AF2+WYFFOS AG, see
\citet{dominguez2014}. The mathematics behind using SVD for least-squares
fitting is discussed by, e.g., \citet*{arun1987}. While mathematically
correct, this approach showed undesirable behaviour in several practical
situations, including sensitivity to noise correlations, poor conditioning for
certain fibre configurations, and occasional slow convergence. We have since
replaced this procedure with a more robust method based on linearized least
squares with iterative refinement, as described below. The new algorithm
provides a fast and highly stable solution for determining the telescope
offsets. It matches or exceeds the accuracy of the earlier SVD-based non-linear
solver in all tested configurations and behaves substantially better in
pathological or low-SNR cases. This approach has now fully replaced the original
method in the WEAVE acquisition and guiding pipeline.
\balance

\subsection*{Linearized formulation}
For a rigid transformation between point pairs $(x_{1,i},y_{1,i})$ and
$(x_{2,i},y_{2,i})$, the exact relation is
\begin{align}
x_{2,i} &= x_{1,i}\cos\phi - y_{1,i}\sin\phi + t_x\\
y_{2,i} &= x_{1,i}\sin\phi + y_{1,i}\cos\phi + t_y,
\end{align}
\noindent where $\phi$ is the rotation angle and $(t_x,t_y)$ the translation.

For sufficiently small $\phi$, we may linearize using $\sin\phi\approx\phi$ and
$\cos\phi\approx 1$, giving
\begin{align}
x_{2,i} &\approx x_{1,i} - \phi y_{1,i} + t_x\\
y_{2,i} &\approx y_{1,i} + \phi x_{1,i} + t_y,
\end{align}
\noindent a formulation that is linear in the unknown parameters
$(\phi, t_x, t_y)$.

Constructing the normal equations yields a $3\times 3$ linear system, which we
solve analytically without SVD. Measurement uncertainties, if supplied, are
incorporated naturally through weighted least squares, and the covariance matrix
of the fitted parameters is obtained directly from the inverted normal matrix.

\subsection*{Iterative refinement}
Although the approximation above is formally valid only for small angles, one or
two iterations suffice to achieve full non-linear accuracy even for several
degrees of rotation. After each iteration, the estimated transformation is
applied to the input points, reducing the remaining non-linear terms, and the
linear fit is repeated. In practice, we found that two iterations always
converge.

This process is equivalent to a Gauss--Newton refinement of the full non-linear
least-squares problem, but with improved conditioning and without the need for
SVD.

\subsection*{Performance and robustness}
This method offers several advantages over the previous SVD solver:

\begin{description}
\item \textit{Numerical stability:} the analytic $3\times 3$ inversion is
extremely well conditioned. It is notably more robust than solving the larger
SVD system, especially for fibre configurations with geometric symmetries or
limited leverage in rotation.
\item \textit{Noise behaviour:} the linearized system avoids cancellations in
the non-linear partial derivatives that occasionally caused poor gradient
behaviour in the previous method. In several test cases the new solver achieved
lower residuals than the SVD fit.
\item \textit{Speed and determinism:} solving a $3\times 3$ system analytically
is computationally negligible, allowing the fit to be repeated several times per
frame with virtually no cost.
\item \textit{Weighted fitting and error propagation:} the linear normal matrix
formulation allows straightforward incorporation of per-axis uncertainties and
yields parameter covariances directly.
\item \textit{Outlier rejection:} an optional clipping stage removes points with
large residuals, significantly improving robustness in the presence of
miscentred, faint, or contaminated guide fibre images.
\end{description}


\vspace*{-3mm} 

\bsp
\end{document}